\documentclass[a4paper,fleqn,usenatbib]{mnras}

\usepackage[T1]{fontenc}
\usepackage{ae,aecompl}

\usepackage{graphicx}	
\usepackage{amsmath}	
\usepackage{amssymb}	
\usepackage{bm}
\usepackage{graphics}
\usepackage{xcolor}
\usepackage{newtxtext} 

\newcommand\rd{{\mathrm{d}}}

\title[Non-linear $P(k)$ in $f(R)$ gravity]{Non-linear matter power spectrum without screening dynamics modelling in $f(R)$ gravity}

\author[C.-Z. Ruan, T.-J. Zhang and B. Hu]{
Cheng-Zong Ruan$^{1,2}$,
Tong-Jie Zhang$^{1}$
and Bin Hu$^{1}$\thanks{E-mail: bhu@bnu.edu.cn}  
\\
$^{1}$Department of Astronomy, Beijing Normal University, Beijing 100875, China\\
$^{2}$Institute for Computational Cosmology, Department of Physics, 
University of Durham, South Road, Durham, DH1 3LE, U.K.
}

\date{Accepted XXX. Received YYY; in original form \today}

\pubyear{2019}

\begin{document}
\label{firstpage}
\pagerange{\pageref{firstpage}--\pageref{lastpage}}
\maketitle

\begin{abstract}
Halo model is a physically intuitive method for modelling the non-linear power spectrum, especially for the alternatives to the standard $\Lambda$CDM models. 
In this paper, we exam the Sheth-Tormen barrier formula adopted in the previous \texttt{CHAM} method \citep{2018MNRAS.476L..65H}. As an example, we model the ellipsoidal collapse of top-hat dark matter haloes in $f(R)$ gravity. 
A good agreement between Sheth-Tormen formula and our result is achieved. The relative difference in the ellipsoidal collapse barrier is less than or equal to $1.6\%$. 
Furthermore, we verify that, for F4 and F5 cases of Hu-Sawicki $f(R)$ gravity, the screening mechanism do not play a crucial role in the non-linear power spectrum modelling up to $k\sim1[h/{\rm Mpc}]$. 
We compare two versions of modified gravity modelling, namely with/without screening. We find that by treating the effective Newton constant as constant number ($G_{\rm eff}=4/3G_N$) is acceptable.  
The scale dependence of the gravitational coupling is sub-relevant. The resulting spectra in F4 and F5, are in $0.1\%$ agreement with the previous \texttt{CHAM} results. The published code is accelerated significantly. 
Finally, we compare our halo model prediction with N-body simulation. We find that the general spectrum profile agree, qualitatively. However, via the halo model approach, 
there exists a systematic under-estimation of the matter power spectrum in the co-moving wavenumber range between $0.3 h/{\rm Mpc}$ and $3 h/{\rm Mpc}$.
These scales are overlapping with the transition scales from two halo term dominated regimes to those of one halo term dominated. 
\end{abstract}




\section{Introduction}
\label{sec:intro}

Non-linear matter power spectrum carries fruitful cosmological information. With the up-coming galaxy surveys, such as Euclid~\footnote{\url{http://sci.esa.int/euclid}}, LSST~\footnote{\url{http://www.lsst.org}}, WFIRST~\footnote{\url{https://wfirst.gsfc.nasa.gov}}, DESI~\footnote{\url{https://www.desi.lbl.gov}}, J-PAS~\footnote{\url{http://www.j-pas.org/wiki/index.php/Main_Page}}, we are aiming to measure the matter power spectrum up to $1\%$ accuracy in the range  from $0.1$ to $10~{\rm Mpc}/h$. Before going to the non-linear part, let us firstly briefly review the status of linear power spectrum modelling for the non-standard cosmologies. This is because the linear spectrum is an essential input for the non-linear part computation.  

For the non-standard cosmologies, we have a few linear Einstein-Boltzmann codes~\footnote{These are all patches to the standard solver, such as CAMB \citep{Lewis:1999bs} and CLASS \citep{Blas:2011rf}.} publicly available on the market, such as \texttt{MGCAMB}~\footnote{\url{https://github.com/sfu-cosmo/MGCAMB}}\citep{Zhao:2008bn,Hojjati:2011ix,Zucca:2019xhg}, \texttt{ISiTGR}~\footnote{\url{https://www.utdallas.edu/~jnd041000/isitgr/}}\citep{Dossett:2011tn,Dossett:2012kd}, \texttt{EFTCAMB}~\footnote{\url{http://eftcamb.org}}\citep{2014PhRvD..89j3530H,2014PhRvD..90d3513R}, \texttt{hi\textunderscore class}~\footnote{\url{https://miguelzuma.github.io/hi_class_public/}}\citep{Zumalacarregui:2016pph}, \emph{etc}. These non-standard Einstein-Boltzmann solvers can be classified into two categories, namely bottom-up and top-down method. The formers are more phenomenologically inspired, such as \texttt{MGCAMB} and \texttt{ISiTGR}. They are built upon the phenomenological parametrizations of non-relativistic gravitational constant ($G_{\rm matter}$) and the relativistic gravitational constant ($G_{\rm light}$)~\footnote{They can also be expressed in term of other related quantities, such as $\mu$, $\gamma$ or $\Sigma$ functions.}. The latter, such as \texttt{EFTCAMB} and \texttt{hi\textunderscore class}, are derived from the first principle point of view, such as the effective field theory of dark energy \citep{Gubitosi:2012hu,Bloomfield:2012ff,Piazza:2013coa}.   

Both of these two philosophies have their advantages and dis-advantages. For the bottom-up method, they are more easily portable among different kinds of surveys, covered from CMB to BAO/RSD observations. It asks for solving much less differential and algebraic equations. However, their dis-advantage is also obvious. These parametrizations are limited to the linear dynamics. The non-linear counter part modelings have some fundamental difficulties, except for some very well-studied theories, such as $f(R)$ gravity \cite{Zhao:2013dza}. This is because, for the current non-linear power spectrum modelling, we can not avoid the calibration from N-body simulation. And, for the N-body simulation of non-standard cosmologies, almost all the algorithms are based on the extra scalar field dynamics modelling. These are completely different modelling languages with respect to the gravitational constant parametrization. Hence, for the moment, one need to cut off all the non-linear data when we adopt the bottom-up method, as an example shown in \cite{Zucca:2019xhg}. Another drawback of the bottom-up method is that some of the parameter space (even they are more favored by the data) are theoretical forbidden \citep{Peirone:2017ywi,Espejo:2018hxa,Frusciante:2018vht}. Thus, this may make our parameter estimation end up in the physically unviable regime. 

The top-down method is anchored to few physical assumptions, such as space-time symmetry arguments. Then, the dynamical system is derived from cosmological linear perturbation theory and dynamical instability analysis (see \cite{Frusciante:2019xia} for view). The state-of-the-art of this method (including code comparison) is nicely summarised in \cite{Bellini:2017avd}. Another merit of this method is that it \emph{can be naturally exported into N-body simulations}. Both linear and non-linear modelling of the extra scalar field dynamics are based on the field theory approach. As demonstrated in our previous work \citep{2018MNRAS.476L..65H}, an accurate linear power spectrum input is essential to the non-linear spectrum calculation. However, the drawback of this method is that, compared with bottom-up method, it is numerically demanding. Hence, it is not easily being transported from one likelihood code to another.   

The non-linear power spectrum modelling methods can be classified into three categories, namely N-body simulation, higher order perturbation theory as well as halo model. Among them, N-body simulation is the most well developed method for non-standard cosmologies, see \cite{Winther:2015wla} for review. The resulting fractional deviation of the matter power spectrum from $\Lambda$CDM agrees to better than $1\%$ up to $k\leq 5-10h {\rm Mpc}^{-1}$ and redshift $z\leq3$ between the different codes for testing examples, such as $f(R)$ gravity, DGP, Symmetron models. 
As for the higher order perturbation theory approach, there exist some comparison of different perturbation theory predictions in the non-standard cosmologies, for example \cite{Valogiannis:2019xed}. Besides these, there are some on-going project on extending \texttt{Pinocchio} algorithm \cite{Monaco:2001jg,Taffoni:2001jh,Monaco:2001jf,Monaco:2013qta} to non-standard cosmologies. Although compared with simulation the semi-analytic halo model (see \cite{Cooray:2002dia} for review) is less accurate, its efficiency is far better than all the other methods. Plus the fact that the current observational data scatters still dominate the error budget. These two aspects inspire us that halo model can be a suitable method for exploring the non-linearities in a wide range of model space. There have already been some studies \citep{Schmidt:2008tn,Lombriser:2016zfz,Lombriser:2013wta,2014JCAP...03..021L,2012MNRAS.421.1431L,2012MNRAS.425..730L,Kopp:2013lea,Achitouv:2015yha} of halo model in the literature based on the spherical/ellipsoidal halo collapse assumptions. In the previous work \citep{2018MNRAS.476L..65H}, we proposed the screened halo model (\texttt{CHAM}) method for the non-linear power spectrum modelling in the alternatives to the standard $\Lambda$CDM scenario. Besides, there also exist some hybrid methods combining higher order perturbation theory with simulations, such as \texttt{COLA} \cite{Tassev:2013pn} and its modified version \cite{Valogiannis:2016ane,Winther:2017jof}. Furthermore, the recent progresses in the emulator \cite{Winther:2019mus} and reaction method \cite{Cataneo:2018cic} predict that we are able to approaching $1\%$ level of modelling the non-linear power spectrum for the generic dark energy/modified gravity models. 

Following our previous work \citep{2018MNRAS.476L..65H}, in this paper we are aiming to validate one of the essential assumption, namely the Sheth-Tormen barrier formula, by modelling the ellipsoidal collapse of top-hat dark matter haloes in $f(R)$ gravity. Throughout this paper, we use the natural unit $c = 1$, where $c$ is the speed of light. An overbar such as $\bar{\rho}_{\mathrm{m}}$ denotes the background value, and a subscript $ _0$ such as $\Omega_{\mathrm{m0}}$ denotes the present value. Primes denote derivatives with respect to $\ln a$, e.g., $D' \equiv \rd D / \rd \ln a$.

The layout of this paper is as follows. 
In Section~\ref{sec:fRgravity}, we briefly review the $f(R)$ gravity theory used in this work. 
In Section~\ref{sec:nonlineardyn}, we present the modeling of the top-hat dark matter halo collapse, in both GR and $f(R)$ gravity. We show the calculation of the collapse barrier, which is a crucial ingredient of the excursion set theory.
Section~\ref{sec:excursion} describes the traditional excursion set theory and the halo model.
Our conclusions are summarized in Section~\ref{sec:summary}.

\section{$f(R)$ gravity}
\label{sec:fRgravity}
In $f(R)$ gravity, the Einstein-Hilbert action is supplemented with a function of the Ricci scalar $R$
\begin{align}
    S = \frac{1}{2\kappa^2} \int \rd^4 x \sqrt{-g} \left[ R + f(R) \right] + S_{\mathrm{m}} (\psi_{\mathrm{m}}; g_{\mu \nu}) \ , 
\end{align}
where $\kappa^2 \equiv 8 \pi G$, $g$ is the determinant of the metric $g_{\mu \nu}$, $S_{\mathrm{m}}$ is the matter action with matter fields $\psi_{\mathrm{m}}$. The modified Einstein equation is derived by varying this action with respect to $g_{\mu \nu}$
\begin{align}
    G_{\mu \nu} + f_R R_{\mu \nu} - \left( \frac{f}{2} - \square f_R \right) g_{\mu \nu} - \nabla_\mu \nabla_\nu f_R = \kappa^2 T_{\mu \nu} \ .
\end{align}
The scalaron $f_R \equiv \rd f / \rd R$ is a new scalar degree of freedom in $f(R)$ gravity. The trace of the modified Einstein equation is the equation of motion for the scalar field 
\begin{align}
    \square f_R = \frac{\partial V_{\mathrm{eff}}}{\partial f_R} \ , \label{equ:eom}
\end{align}
with the effective potential defined as 
\begin{align}
    \frac{\partial V_{\mathrm{eff}}}{\partial f_R} \equiv \frac{1}{3} \left[ R - f_R R + 2f - \kappa^2 (\rho - 3p) \right] \ .
\end{align}
The curvature of this potential, which can be regarded as the effective mass of the field $f_R$, is given by 
\begin{align}
    m_{f_R}^2 = \frac{\partial^2 V_{\mathrm{eff}}}{\partial f_R^2} = \frac{1}{3} \left( \frac{1 + f_R}{f_{RR}} - R \right) \ , 
\end{align}
where $f_{RR} \equiv \rd^2 f / \rd R^2$. Hereafter, we adopt the most well-studied example of $f(R)$ gravity, Hu-Sawicki $f(R)$ gravity model \citep{2007PhRvD..76f4004H}, which can satisfy the background $\Lambda$CDM expansion history and evade the Solar system tests. 
The formula of the extra gravity term can be written as
\begin{align}
    f(R) = -2\Lambda - \bar{f}_{R0} \frac{\bar{R}_0^2}{R} \ , \label{equ:fRLambda}
\end{align}
where $\Lambda$ is an effective cosmological constant driving the accelerating cosmic expansion. 
In the limit of $|f_{R0}| \ll 1$, the background expansion history is almost the same as $\Lambda$CDM model \citep{2007PhRvD..76f4004H,2008PhRvD..78l3524O}, in which the background Ricci scalar can be approximated as 
\begin{align}
    \bar{R} \approx 3H_0^2 \left[ \Omega_{\mathrm{m0}}(1+z)^3 + 4\Omega_{\Lambda 0} \right] \ , \label{equ:Rbar}
\end{align}
where the density fraction is given by $\Omega_{i0} \equiv 8\pi G \bar{\rho}_{i0} / (3H_0^2), i = \{\mathrm{m}, \Lambda \}$.

\subsection{Cosmic linear perturbation regime}
In scalar-tensor theories such as $f(R)$ gravity, the linear growth function of matter fluctuations $D (a, k)$ becomes scale dependent. The linear growth function $D$ is defined as
\begin{align}
    D(a, k;a_{\mathrm{init}}) \equiv \frac{\delta_{\mathrm{m}} (a, k) }{\delta_{\mathrm{m}} (a_{\mathrm{init}}, k)} \ ,
\end{align}
where $\delta_{\mathrm{m}} (a, \bm{x}) \equiv \rho_{\mathrm{m}} (a, \bm{x})/\bar{\rho}_{\mathrm{m}}(a) - 1$ is the matter overdensity and $\delta_{\mathrm{m}} (a, k)$ is the Fourier transform.
In the quasi-static limit, the evolution equation of the growth function is   (see, e.g., \citet{2014AnP...526..259L})
\begin{align}
    D'' + \left[ 2 - \frac{3}{2} \Omega_{\mathrm{m}}(a) \right] D' - \frac{3}{2} \mu(a, k)\, \Omega_{\mathrm{m}}(a) \, D \approx 0 \ , \label{equ:Ddiff}
\end{align}
where $\Omega_{\mathrm{m}}(a) \equiv H_0^2 \Omega_{\mathrm{m}0} a^{-3} / H^2(a)$. $\mu (a, k)$ is the modification in the Poisson equation due to the scalar field, which takes the form 
\begin{align}
    \mu (a, k) \approx 1 + \frac{1}{3} \frac{k^2}{a^2 \bar{m}^2 + k^2} \ ,
\end{align}
with $\bar{m}^2 \approx \left[ 3 f_{RR} (R=\bar{R}) \right]^{-1}$ is the mass of the scalaron evaluated at the background. Combining the $f(R)$ function form in Equation~\eqref{equ:fRLambda} with the expression of $\bar{R}$ (Equation~\eqref{equ:Rbar}), we have
\begin{align}
    \bar{m} &= \frac{\left( \Omega_{\mathrm{m0}}a^{-3} + 4\Omega_{\Lambda 0} \right)^{3/2} \, }{3\times 10^3 \sqrt{2|f_{R0}|} (\Omega_{\mathrm{m0}} + 4\Omega_{\Lambda 0}) } h\,\mathrm{Mpc}^{-1} \ .
\end{align}


The above algorithm capture the major feature of linear matter growth in modified gravity, namely the scale dependence. For an accurate calculation, we need to take into account the other ingredients, such as baryon and neutrino.  
For this purpose, we utilise the more sophisticated linear Einstein-Boltzmann solver~\texttt{EFTCAMB} \cite{2014PhRvD..89j3530H,2014PhRvD..90d3513R}. 
Specifically, for $f(R)$ gravity we use the code developed in \cite{Hu:2016zrh}. 

\subsection{Nonlinear regime}
\label{subsec:chameleon}
\citet{PhysRevD.69.044026} derived an estimation of the radial profile of the scalar field $\varphi (r) \equiv f_R (r)$, in a spherically symmetric top-hat overdensity of (physical) radius $\xi_{\mathrm{TH}}$ with constant inner and outer matter density $\rho_{\mathrm{in}}$ and $\rho_{\mathrm{out}}$, respectively. 
The solutions of the scalar field, $\varphi (r)$, minimize the effective potential $V_{\mathrm{eff}} (\varphi)$ in the equation of motion \eqref{equ:eom}.
If $\rho_{\mathrm{in}} = \rho_{\mathrm{out}}$, then $\varphi$ will be constant in the whole space. 
When $\rho_{\mathrm{in}} \neq \rho_{\mathrm{out}}$, if we go towards the center of the sphere from outside, the field value will settle from constant $\varphi_{\mathrm{out}}$ (at far outside) to another constant $\varphi_{\mathrm{in}}$, as long as the difference between the two values are not too large. 
\citet{PhysRevD.69.044026} find that the radial profile $\varphi(r)$ in the \textit{thin-shell regime} is 
\begin{align}
    \varphi(r) \approx \begin{cases} \displaystyle
        \varphi_{\mathrm{in}} \ , & r \le \xi_0 \\
        \displaystyle \varphi_{\mathrm{in}} + \frac{\kappa \beta}{3} \rho_{\mathrm{in}} \left( \frac{r^2}{2} + \frac{\xi_0^3}{r} - \frac{3}{2} \xi_0^2 \right)  \ , &  \xi_0 < r \le \xi_{\mathrm{TH}} \\
        \displaystyle \varphi_{\mathrm{out}} - \frac{\Delta \xi}{\xi_{\mathrm{TH}}} \frac{\sqrt{\kappa} \gamma \rho_{\mathrm{in}} \xi_{\mathrm{TH}}^3}{r} e^{-m_{\mathrm{out}}(r - \xi_{\mathrm{TH}})} \ , & r > \xi_{\mathrm{TH}}
    \end{cases} \ ,
\end{align}
where $\beta = -1/\sqrt{6}$ for $f(R)$ gravity; $\Delta \xi \equiv \xi_{\mathrm{TH}} - \xi_0 \ll 1$ is the thickness of the thin-shell, and $m_{\mathrm{out}} \equiv \rd^2 V_{\mathrm{eff}} (\varphi_{\mathrm{out}}) / \rd \varphi^2$ is the effective mass of the outside field.
The distance needed for $\varphi$ to settle from $\varphi_{\mathrm{out}}$ to $\varphi_{\mathrm{in}}$ is \citep{2012MNRAS.421.1431L,2013PhRvD..87l3511L,2014JCAP...03..021L}
\begin{align}
    \frac{\Delta \xi}{\xi_{\mathrm{TH}}} &\approx \frac{|f_{R0}| a^3}{\Omega_{\mathrm{m0}} \tilde{\rho}_{\mathrm{in}} (H_0 \xi_{\mathrm{TH}})^2 } \times  \bigg[ \left( \frac{1 + 4\Omega_{\mathrm{\Lambda 0}}/\Omega_{\mathrm{\mathrm{m} 0}}}{\tilde{\rho}_{\mathrm{out}} a^{-3} + 4\Omega_{\mathrm{\Lambda 0}}/\Omega_{\mathrm{\mathrm{m} 0}} } \right)^2  \notag \\
    &\phantom{=} - \left( \frac{1 + 4\Omega_{\mathrm{\Lambda 0}}/\Omega_{\mathrm{\mathrm{m} 0}}}{\tilde{\rho}_{\mathrm{in}} a^{-3} + 4\Omega_{\mathrm{\Lambda 0}}/\Omega_{\mathrm{\mathrm{m} 0}} } \right)^2 \bigg] \ , \label{equ:deltaxi}
\end{align}
where $\tilde{\rho}_{\mathrm{in/out}} \equiv \rho_{\mathrm{m,in/out}} / \bar{\rho}_{\mathrm{m}}$. The enhancement of gravity (the fifth force) due to the extra scalar field for a unity test particle at $r = \xi_{\mathrm{TH}}$ is 
\begin{align}
    \mathcal{F} \frac{GM_{\mathrm{TH}}}{\xi_{\mathrm{TH}}^2} &\equiv \kappa \beta |\nabla \varphi|_{r=\xi_{\mathrm{TH}}} \approx 2\beta^2 \frac{GM_{\mathrm{TH}}}{\xi_{\mathrm{TH}}^2} \left[ 1 - \left( \frac{\xi_0}{\xi_{\mathrm{TH}}} \right)^3 \right] \\
    &= 2\beta^2 \frac{GM_{\mathrm{TH}}}{\xi_{\mathrm{TH}}^2} \left[ 3 \frac{\Delta \xi}{\xi_{\mathrm{TH}}} - 3 \left( \frac{\Delta \xi}{\xi_{\mathrm{TH}}} \right)^2 + \left( \frac{\Delta \xi}{\xi_{\mathrm{TH}}} \right)^3 \right] \ . \label{equ:fff}
\end{align}
Since $\xi_{\mathrm{TH}} \ge \xi_0 > 0$, the ratio $\Delta \xi / \xi_{\mathrm{TH}} \in [0, 1]$, which means the enhancement of gravity $\mathcal{F} \in [0, 1/3]$. For a top-hat overdensity, the last equation provides an interpolation between the screened and un-screened regime. We shall follow \citet{2013PhRvD..87l3511L} and use Equation~\eqref{equ:fff} as the force enhancement when studying the spherical and ellipsoidal collapse model:
\begin{align}
    \mathcal{F} = \frac{1}{3} \mathrm{min}\, \left\{ \left[ 3 \frac{\Delta \xi}{\xi_{\mathrm{TH}}} - 3 \left( \frac{\Delta \xi}{\xi_{\mathrm{TH}}} \right)^2 + \left( \frac{\Delta \xi}{\xi_{\mathrm{TH}}} \right)^3 \right], 1 \right\}\ . \label{equ:fff2}
\end{align}

\section{Collapsing process}
\label{sec:nonlineardyn}
In this section, we first review the spherical collapsing in $f(R)$ gravity and ellipsoidal collapsing in GR. Then, we solve the ellipsoidal collapsing process in $f(R)$ gravity. 

\subsection{Spherical collapse in $f(R)$ gravity}
We study the formation of dark matter halos in $f(R)$ gravity using both the spherical and ellipsoidal collapse models. We approximate the dark matter halo by a top-hat over density within the initial comoving radius $R_{\mathrm{init}}$. 
Afterward, the local density, $\rho_{\mathrm{m}}(a)$, changes due to the physical radius, $\xi (a)$, changes with time. 
In the initial matter-dominated era, $\xi (a_{\mathrm{init}}) =a_{\mathrm{init}} R_{\mathrm{init}}$. We define the dimensionless comoving radius $y(a)$ as 
\begin{align}
    y(a) \equiv \frac{\xi(a) / a}{R_{\mathrm{init}}} \ ,
\end{align}
so that $y(a_{\mathrm{init}}) = 1$. The conservation of mass in the top-hat region implies $\bar{\rho}_{\mathrm{m,init}} a_{\mathrm{init}}^3 R^3_{\mathrm{init}} = \rho_{\mathrm{m}} \xi^3(a)$, thus $\tilde{\rho} \equiv \rho_{\mathrm{m}} / \bar{\rho}_{\mathrm{m}} = y^{-3} (a)$. 

The spherical collapse equation in $f(R)$ gravity is given by \citep{2009PhRvD..79h3518S,2012MNRAS.421.1431L} 
\begin{align}
    \frac{1}{\xi} \frac{\rd^2 \xi}{\rd t^2} = -\frac{\kappa^2}{6} (\bar{\rho}_{\mathrm{m}} - 2 \bar{\rho}_{\Lambda}) - \frac{\kappa^2}{6} (1 + \mathcal{F}) \delta \rho_{\mathrm{m}} \ .
\end{align}

Replacing $\xi(a)$ with $y(a)$ and the time variable $t$ with $\ln a$ yields 
\begin{align}
    y''_{\mathrm{h}} &+ \left[2 - \frac{3}{2} \Omega_{\mathrm{m}} (a) \right] y'_{\mathrm{h}} \notag \\
    \quad &+ \frac{1}{2} \Omega_{\mathrm{m}}(a) \left[1 + \mathcal{F}(a; y_{\mathrm{h}}, y_{\mathrm{env}}) \right] (y_{\mathrm{h}}^{-3} - 1) y_{\mathrm{h}} = 0 \ , \label{equ:SCMG}\\
    y''_{\mathrm{env}} &+ \left[2 - \frac{3}{2} \Omega_{\mathrm{m}} (a) \right]  y'_{\mathrm{env}} + \frac{1}{2} \Omega_{\mathrm{m}} (a) (y_{\mathrm{env}}^{-3} - 1) y_{\mathrm{env}} = 0 \ , \label{equ:yenv}
\end{align}
where $\Omega_{\mathrm{m}} (a) = \frac{\Omega_{\mathrm{m0}} a^{-3}}{H^2/H_0^2} = \frac{\Omega_{\mathrm{m0}} a^{-3}}{\Omega_{\mathrm{m0}} a^{-3} + \Omega_{\Lambda 0}}, \Omega_{\mathrm{m0}} + \Omega_{\Lambda0} = 1$. The subscripts $_\mathrm{h}$ and $_\mathrm{env}$ denote the inner and outer overdensities, i.e., the halo and its local environment, respectively. The modification of gravity $\mathcal{F} (\Delta \xi / \xi)$ is given by the thin-shell approximation in Section~\ref{subsec:chameleon}. According to Equation~\eqref{equ:deltaxi}, the thickness of the thin-shell is 
\begin{align}
    \frac{\Delta \xi}{\xi}(a) &= \frac{|f_{\mathrm{R0}}|}{\Omega_{\mathrm{\mathrm{m0}}}}  \left(\frac{c}{H_0 R_{\mathrm{init}}} \right)^2 a^7 y_{\mathrm{h}} \bigg[ \left( \frac{1 + 4 \Omega_{\mathrm{\Lambda 0}}/\Omega_{\mathrm{\mathrm{m0}}}}{y_{\mathrm{env}}^{-3} + 4 (\Omega_{\Lambda0}/\Omega_{\mathrm{\mathrm{m0}}}) a^3} \right)^2 \notag \\
    & \quad - \left( \frac{1 + 4 \Omega_{\mathrm{\Lambda 0}}/\Omega_{\mathrm{\mathrm{m0}}}}{y_{\mathrm{h}}^{-3} + 4 (\Omega_{\Lambda0}/\Omega_{\mathrm{\mathrm{m0}}}) a^3} \right)^2 \bigg]  \ ,\label{equ:DRdivRTH}
\end{align}
and the factor $\mathcal{F} (\Delta \xi / \xi)$ is given by Equation~\eqref{equ:fff2}. We have assumed that the environment follows $\Lambda$CDM evolution, i.e., the modification of gravity $\mathcal{F} = 0$, in Equation~\eqref{equ:yenv}.

Equations~\eqref{equ:SCMG} and \eqref{equ:yenv} form a system of coupled differential equations for $y_\mathrm{h} (a)$ and $y_{\mathrm{env}} (a)$. To solve these equations, we set the initial conditions at $a_{\mathrm{init}} \ll 1$ in the matter-dominated regime:
\begin{align}
    y_{\mathrm{h/env,init}} = 1 - \frac{\delta_{\mathrm{h/env,init}}}{3} \ , \quad y'_{\mathrm{h/env,init}} = -\frac{\delta_{\mathrm{h/env,i}}}{3} \ .
\end{align}

For a fixed initial time $a_{\mathrm{init}}$, we adjust the initial overdensity $\delta_{\mathrm{h,init}}$ so that $y_{\mathrm{h}} (a_0) = 0$, i.e., the top-hat halo collapses at present time. The extrapolated linear spherical critical density (also called collapse barrier) $\delta_{\mathrm{sc}}^{f(R)}$ used in the excursion set formalism is defined by 
\begin{align}
    \delta_{\mathrm{sc}}^{f(R)} &\equiv D(a_0, k_{\mathrm{h}};a_{\mathrm{init}}) \delta_{\mathrm{h,init}} \ , \label{equ:deltascfr} \\
    \delta_{\mathrm{env}} &\equiv D_{\Lambda\mathrm{CDM}} (a_0;a_{\mathrm{init}}) \delta_{\mathrm{env,init}} \ ,
\end{align}
where $D(a, k)$ is the $f(R)$ gravity linear growth function solved from Equation~\eqref{equ:Ddiff}. For a top-hat halo with mass $M_{\mathrm{h}} = \frac{4\pi}{3} (R_{\mathrm{init}} a_{\mathrm{init}})^3 \bar{\rho}_{\mathrm{m,init}}$, its corresponding wavenumber $k_{\mathrm{h}} \equiv 1 / R_{\mathrm{init}}$ is 
\begin{align}
    \frac{c}{H_0 R_{\mathrm{init}}} = 3\times 10^3 \times \left( \frac{1.12\pi}{3} \right)^{1/3} \Omega_{\mathrm{m0}}^{1/3}  \left( \frac{M_{\mathrm{h}}}{10^{12}\, h^{-1}\, M_\odot} \right)^{-1/3}  \ . \label{equ:kheqn}
\end{align}

The extrapolated linear value for environment $\delta_{\mathrm{env}}$ is defined by $\Lambda$CDM linear growth function (see, e.g., \citet{2003moco.book.....D})
\begin{align}
    D_{\Lambda \mathrm{CDM}} (a) = \frac{5\Omega_{\mathrm{m0}}}{2} \frac{H(a)}{H_0} \int_0^a \frac{\rd a'}{\big[a' H(a') / H_0\big]^3} \ .
\end{align}
As we have discussed above, the spherical collapse barrier in $f(R)$ gravity depends on both the halo mass, $M_{\mathrm{h}}$, and environment overdensity, $\delta_{\mathrm{env}}$ \footnote{The initial overdensity $\delta_{\mathrm{h,init}}$ is restricted with condition $y_{\mathrm{h}}(a_0) = 0$, so that it is not a free variable. }. 


\subsection{Ellipsoidal collapse in GR}\label{sec:GREC}
The spherical symmetry is an over-simplification of the collapsing process. 
\citet{1970Ap......6..320D} has shown that a initially spherical overdensity embedded in a Gaussian perturbation field would evolves into triaxial ellipsoid, approximately. 
The three main axes of the ellipsoid are aligned with three eigen vectors of the so-called deformation tensor $\propto \nabla_i \nabla_j \Phi$, where $\Phi$ is the gravitational potential perturbation \citep{2010gfe..book.....M}. 
Thus, the collapse of a homogeneous ellipsoid should provide a better description of halo formation and collapse barrier.

The dynamics of the ellipsoid is set by the potential perturbations due to the matter interior and exterior to the ellipsoid, respectively. 
The Euler equation of a fluid element at the comoving coordinates $\bm{x}$ inside the ellipsoid is 
\begin{align}
    \frac{\rd \bm{v}}{\rd t} = -\frac{1}{a} \nabla \Phi (\bm{x}) \ , \label{equ:eulereqn}
\end{align}
where $\bm{v}$ is the peculiar velocity, and the gravitational potential perturbation $\Phi$ obeys the Poisson equation
\begin{align}
    \nabla^2 \Phi = 4\pi G \bar{\rho}_{\mathrm{m}}(a)\, a^2 \, \Delta(a) \ , \label{equ:poieqn}
\end{align}
with $\Delta (a) \equiv \big[\rho_{\mathrm{m}}(a) - \bar{\rho}_{\mathrm{m}}(a)\big] / \bar{\rho}_{\mathrm{m}}(a) \approx \rho_{\mathrm{m}}(a) / \bar{\rho}_{\mathrm{m}}(a)$ is the (non-linear) overdensity of the top-hat ellipsoid.

$\Phi$ can be separated in inside (ellipsoid's self-gravity) and outside two parts, $\Phi = \Phi_{\mathrm{int}} + \Phi_{\mathrm{out}}$. The inner part of gravitational potential from the homogeneous ellipsoid has analytical form. As for the outside part, it can be neglected in the deep non-linear regime since the density contrast of the ellipsoid is high enough to dominate the dynamics. However, in order to give a correct initial condition for the nonlinear collapsing process, we can not completely ignore the external potential. 
It has been proven \citep{2010gfe..book.....M} that $\Phi_{\mathrm{out}}$ can be approximated by linear perturbation. Let's consider the ellipsoidal originated from a spherical overdense regime with initial comoving radius $R_{\mathrm{init}}$. According to the Zel'dovich approximation \citep{1970A&A.....5...84Z}, the sphere evolves into an ellipsoid with principal axes $X_i(a) = R_{\mathrm{init}} \big[ 1 - \lambda_i D(a) / D(a_{\mathrm{init}}) \big] = \big[ 1 - \lambda_i D(a) / a_{\mathrm{init}} \big]$ in the matter dominated linear regime, \footnote{The terms containing linear growth function $D(a)$ differs by a constant factor $D(a_{\mathrm{init}})$ with the original form in \citet{2010gfe..book.....M}, for the different normalization of $D(a)$.} where $\lambda_i (i=1,2,3,)$ are the eigenvalues of the deformation tensor, $\nabla_i \nabla_j \Phi_{\mathrm{init}} / (4\pi G \bar{\rho}_{\mathrm{m}} a^3)$. Thus, the principal axes of the ellipsoid are parallel to those of the tidal shear field.

Combining the Euler equation~\eqref{equ:eulereqn}, the Poisson equation~\eqref{equ:poieqn}, the Zel'dovich approximation and the mass convervation of the ellipsoid $M_{\mathrm{h}} = \frac{4\pi}{3} \big[1+\Delta(a)\big] \bar{\rho}_{\mathrm{m}} a^3 X_1 X_2 X_3$, the dynamical equations of the principal axes' comoving length $X_j (t)$ are \citep{2010gfe..book.....M}
\begin{align}
    \frac{\rd^2 X_j}{\rd t^2} + 2\frac{\rd a / \rd t}{a} \frac{\rd X_j}{\rd t} = &-4\pi G \bar{\rho}_{\mathrm{m}}(t)\, X_j \bigg[ \frac{1}{2} \alpha_j(t) \Delta(t) \notag \\
    & + \frac{D(t)}{a_{\mathrm{init}}} \left( \lambda_j - \frac{1}{3} \delta_{\mathrm{init}} \right)\bigg] \ , \label{equ:ECGR1}
\end{align}
where 
\begin{align}
    \alpha_j(t) \equiv X_1 X_2 X_3 \int_0^\infty \rd y (X_j^2 + y)^{-1} \prod_{k=1}^3 (X_k^2 + y)^{-1/2} \ ,
\end{align}
is related to the ellipsoidal geometry. 

Defining the dimensionless comoving length 
\begin{align}
Y_j(t) \equiv X_j(t) / R_{\mathrm{init}} \ ,
\end{align} 
and changing time variable from $t$ to $\ln a$, Equation~\eqref{equ:ECGR1} is expressed as  
\begin{align}
    Y''_j + \left[2 - \frac{3}{2} \Omega_{\mathrm{m}}(a) \right] Y'_j =& -\frac{3}{2} \Omega_{\mathrm{m}}(a)\, Y_j \bigg[ \frac{1}{2} \alpha_j \Delta  \notag  \\
     &+ \frac{D(a)}{a_{\mathrm{init}}} \left( \lambda_j - \frac{1}{3}\delta_{\mathrm{h,init}}   \right) \bigg] \ . \label{equ:ECGR2}
\end{align}

To solve these equations, we set the initial conditions in matter dominated regime $a_{\mathrm{init}} \ll 1$ according to the Zel'dovich approximation, 
\begin{align}
    Y_j(a_{\mathrm{init}}) &= 1 - \lambda_j  \ , \label{eqn:ECic1} \\
    Y'_j (a_{\mathrm{init}}) &= -\lambda_j \ . \label{eqn:ECic2}
\end{align}
Thus, the initial conditions are fully specified by $\{ \lambda_j, j = 1, 2, 3, \ \text{assuming}\ \lambda_1 \ge \lambda_2 \ge \lambda_3 \}$. In practice, they are presented by the initial density of the ellipsoid
\begin{align}
    \delta_{\mathrm{init}} = \lambda_1 + \lambda_2 + \lambda_3 \ ,
\end{align}
the ellipticity in the $(\lambda_1, \lambda_3)$ plane:
\begin{align}
    e \equiv \frac{\lambda_1 - \lambda_3}{2 \delta_{\mathrm{init}}} \ ,
\end{align}
and the oblateness (when $0 \le p \le e$) or prolateness ($0 \ge p \ge -e$) of the ellipsoid:
\begin{align}
    p \equiv \frac{\lambda_1 + \lambda_3 - 2\lambda_2}{2 \delta_{\mathrm{init}}} \ .
\end{align}
Sphere corresponds to $e = p = 0$.

According to Equation~\eqref{equ:ECGR2}, the shortest axis collapses to $Y=0$ first, after which Equation~\eqref{equ:ECGR2} is not valid. To alleviate this problem, it is usually assumed that collapse along each axis is frozen once the axis has shrunk to a freeze-out radius.
The virialization of ellipsoid is identified by the freeze-out of the longest axis, so that virial overdensity (the overdensity at the time of virialization) equals to $179$, which reproduces the spherical collapse result. The ellipsoidal collapse barrier $\delta_{\mathrm{sc}}^{\mathrm{GR}}$ in general relativity depends on ellipticity parameters $e$ and $p$ ($\delta_{\mathrm{init}}$ is not free parameter):
\begin{align}
    \delta_{\mathrm{ec}}^{\mathrm{GR}} = \delta_{\mathrm{ec}}^{\mathrm{GR}} (e, p) \ .
\end{align}

By fitting the values of $\delta_{\mathrm{ec}}^{\mathrm{GR}} (e, p)$ from the ellipsoidal collapse model described above, \citet{2001MNRAS.323....1S} found the ellipsoidal collapse barrier can be approximated by solving
\begin{align}
    \frac{\delta_{\mathrm{ec}}^{\mathrm{GR}} (e, p)}{\delta_{\mathrm{sc}}^{\mathrm{GR}}} \approx 1 + \beta \left\{ 5 (e^2 \pm p^2) \left[ \frac{\delta_{\mathrm{ec}}^{\mathrm{GR}}(e, p) }{\delta_{\mathrm{sc}}^{\mathrm{GR}}} \right]^2 \right\} \ , \label{equ:STori}
\end{align}
where $\beta = 0.47, \gamma = 0.615$ and $\delta_{\mathrm{sc}}^{\mathrm{GR}}$ is the spherical collapse barrier.

\subsection{Ellipsoidal collpase in $f(R)$ gravity}
\label{sec:FREC}

There are research works focusing on the chameleon screening mechanism in non-spherical cases (e.g., \citet{2018JCAP...01..056B}). \citet{2015PhRvD..91f5030B} discussed the full form of ellipsoidal chameleon force. They found that, in extreme situations ($\sim 0.99$ ellipticity), enhancement of the chameleon force would differ by up to $40\%$ for a sphere and an ellipsoid with the same mass. In the following subsection, we will provide our calculation. 

We present a simple ellipsoidal collapse of top-hat overdensity in $f(R)$ gravity, which combines the ingredients of the above two subsections. 
Considering a homogeneous ellipsoid embedded in a larger spherical environment, the fluid element inside the top-hat halo experiences the modified gravity, $G_{\mathrm{eff}} = (1 + \mathcal{F}) G$. 
We approximate this effect of the fifth force as the spherical case discussed in Section~\ref{subsec:chameleon}. That is, replacing the spherical radius $y$ with an `effective' length $(Y_1 Y_2 Y_3)^{1/3}$ in the expression of the thickness of thin-shell Equation~\eqref{equ:DRdivRTH}. 

Similar with Equation~\eqref{equ:DRdivRTH}, the thickness of the thin-shell and the force enhancement are 
\begin{align}
    \frac{\Delta \xi}{\xi} = & \frac{|f_{\mathrm{R0}}| c^2 a^7}{\Omega_{\mathrm{\mathrm{m0}}} (H_0 R_{\mathrm{init}})^2 }  (Y_1 Y_2 Y_3)^{1/3}  \bigg[ \left( \frac{1 + 4 \Omega_{\mathrm{\Lambda 0}}/\Omega_{\mathrm{\mathrm{m0}}}}{y_{\mathrm{env}}^{-3} + 4 (\Omega_{\Lambda0}/\Omega_{\mathrm{\mathrm{m0}}}) a^3} \right)^2 \notag \\
    &- \left( \frac{1 + 4 \Omega_{\mathrm{\Lambda 0}}/\Omega_{\mathrm{\mathrm{m0}}}}{(Y_1 Y_2 Y_3)^{-1} + 4 (\Omega_{\Lambda0}/\Omega_{\mathrm{\mathrm{m0}}}) a^3} \right)^2 \bigg]  \ ,\label{equ:DRdivRTH2} \\
    \mathcal{F} = &\frac{1}{3}\, \mathrm{min} \left[ 3 \frac{\Delta \xi}{\xi} - 3 \left( \frac{\Delta \xi}{\xi} \right)^2 + \left( \frac{\Delta \xi}{\xi} \right)^3, 1 \right]  \ .
\end{align}

The ellipsoidal collapse equations of $Y_j$ (dimensionless comoving length of principle axes of the ellipsoid) in $f(R)$ gravity can be written as  
\begin{align}
    Y''_j + \left[2 - \frac{3}{2} \Omega_{\mathrm{m}}(a) \right] Y'_j = -&\frac{3}{2} \left[ 1+\mathcal{F}(a; Y_j, y_{\mathrm{env}}) \right] \Omega_{\mathrm{m}}(a)\, Y_j \notag  \\
    \times &\left[ \frac{1}{2} \alpha_j \Delta + \frac{D(a)}{a_{\mathrm{init}}} \left( \lambda_j - \frac{1}{3}\delta_{\mathrm{h,init}}   \right) \right] \ . \label{equ:ECMG3}
\end{align}
When we come back to spherical case, i.e., $Y_1 = Y_2 = Y_3 \equiv y_{\mathrm{h}}$, and ignore the tidal force term $\propto (\lambda_j - \frac{1}{3} \delta_{\mathrm{h,init}})$, Equation~\eqref{equ:ECMG3} is consistent with Equation~\eqref{equ:SCMG}, as it is supposed to be. The scale-independent linear growth function in $\Lambda$CDM model $D = D_{\Lambda\mathrm{CDM}}(a)$ is used in Equation~\eqref{equ:ECMG3}, although the more natural choice is to use the growth function in $f(R)$ gravity. We have checked that this approximation has little effect, since in early matter dominated regime all linear growth functions should be proportional to scale factor, and in late time the linear term is unimportant. 

$\Lambda$CDM evolution of the spherical environment is assumed as before
\begin{align}
    y''_{\mathrm{env}} + \left[2 - \frac{3}{2} \Omega_{\mathrm{m}} (a) \right]  y'_{\mathrm{env}} + \frac{1}{2} \Omega_{\mathrm{m}} (a) (y_{\mathrm{env}}^{-3} - 1) y_{\mathrm{env}} = 0 \ , \label{equ:yenv2}
\end{align}
with the same initial conditions 
\begin{align}
    Y_j(a_{\mathrm{init}}) &= 1 - \lambda_j  \ ,  Y'_j (a_{\mathrm{init}}) = -\lambda_j \ , \\
    y_{\mathrm{env,init}} &= 1 - \frac{\delta_{\mathrm{env,init}}}{3} \ , \  y'_{\mathrm{env,init}} = -\frac{\delta_{\mathrm{env,init}}}{3} \ .
\end{align}

\begin{figure}
    \centering 
    \includegraphics[width=\columnwidth]{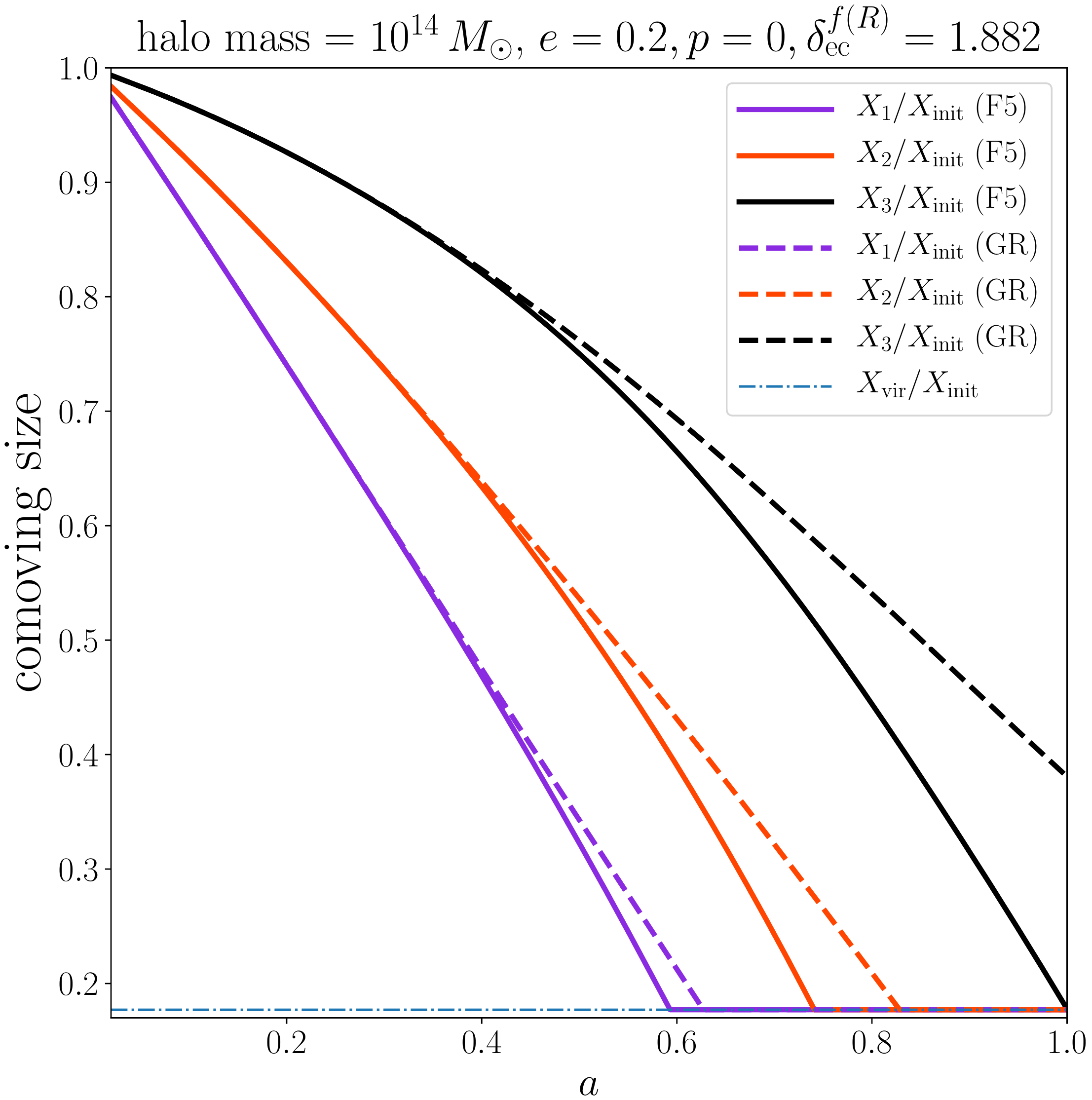}
    \caption{ Ellipsoidal collapse of the top-hat overdensity in $f(R)$ gravity ($|f_{R0}| = 10^{-5}$, F5, solid line) and general relativity (dashed line). The initial conditions are shown in the title of the figure, besides, $\delta_{\mathrm{env}} = 0.8$.  Halos collapse faster in $f(R)$ gravity than in GR, since the gravity is enhanced. The freeze-out mechanism ensures that the overdensity at virialization (defined as when the longest axis is frozen) equals to $179$, which is predicted by spherical collapse. }
    \label{fig:ECplot_FR_GR}
\end{figure}

To fully specify the ellipsoidal collapse process in $f(R)$ gravity, the parameters $M_{\mathrm{h}}$ (or equivalent $R_{\mathrm{init}}$), $\delta_{\mathrm{env}}$, $e$, $p$ along with cosmological parameters such as $\Omega_{\mathrm{m0}}$ should be given. 
For a fixed initial time $a_{\mathrm{init}}$, we adjust the initial overdensity $\delta_{\mathrm{h,init}}$ so that the longest axis of the ellipsoid is frozen at $a = 1$. 
We use the $f(R)$ gravity linear growth function $D(a, k)$ from Equation~\eqref{equ:Ddiff} to extrapolate $\delta_{\mathrm{h,init}}$ to the present time, defining the ellipsoidal collapse barrier
\begin{align}
    \delta_{\mathrm{ec}}^{f(R)} ( \underbrace{M_{\mathrm{h}}, \delta_{\mathrm{env}}}_{\text{MG effect}}, \underbrace{e, p}_{\text{EC effect}}) \equiv D(a=1, k_{\mathrm{h}}; a_{\mathrm{init}}) \delta_{\mathrm{h,init}} \ ,
\end{align}
where $k_{\mathrm{h}} \equiv 1 / R_{\mathrm{init}}$ is given by Equation~\eqref{equ:kheqn}. Figure~\ref{fig:ECplot_FR_GR} shows an example of ellipsoidal collpase model we described above. Given initial conditions $M_{\mathrm{h}} = 10^{14} M_\odot, \delta_{\mathrm{env}} = 0.8, e = 0.2$ and $p = 0$, by adjusting $\delta_{\mathrm{h,init}}$ so that the longest axis (black solid line in Figure~\ref{fig:ECplot_FR_GR}) is frozed at $a=1$.  We find the ellipsoidal collpase barrier in $f(R)$ gravity with $|f_{R0}| = 10^{-5}$, for halo with mass $10^{14} M_\odot$ is $\delta_{\mathrm{ec}}^{f(R)} = 1.882$. 
The values of cosmological parameters follow the $f(R)$ gravity $N$-body simulation presented by \citet{2013MNRAS.428..743L}. They are set to $\Omega_{\mathrm{m0}} = 0.24$ with $\Omega_{\Lambda 0} = 1 - \Omega_{\mathrm{m0}}$, $h = 0.73$ for the dimensionless Hubble constant, $n_s = 0.958$ for the slope of the primordial power spectrum and the power spectrum normalization $\sigma_8 = 0.8$ in $\Lambda$CDM.

\section{Semi-analytical methods for halo mass function and power spectrum}
\label{sec:excursion}
In the following section, we will firstly review the idea of excursion set formalism for halo distribution in mass spectrum. And then, introduce our method. 

\subsection{Excursion set formalism}
The trajectory of the excursion sets are constructed from the filtered linear density field with different smoothing scales. 
A dark matter halo can be formed once the filtered density cross up the critical value on the largest scales \citep{1988MNRAS.235..715E,1989ApJ...340...47C,1991ApJ...379..440B}. 
According to this idea, one can use the statistics of linear perturbation field to infer the (comoving) number density of haloes as a function of mass, i.e., the halo mass function.

Considering the linear perturbation field extrapolated to the phase of non-linear evolution $\delta_{\mathrm{lin}} (\bm{x}, t) = D(t) \delta_{\mathrm{init}} (\bm{x})$, according to the spherical or ellipsoidal collapse model presented in \S~\ref{sec:nonlineardyn}, regions with $\delta_{\mathrm{lin}} (\bm{x}, t) > \delta_c$, or equivalently, $\delta_{\mathrm{init}} (\bm{x}) > \delta_c / D(t) \equiv \delta_c(t)$, have collapsed into dark matter haloes. 
To assign a halo with a mass, \citet{1974ApJ...187..425P} assumed that, the probability that smoothed density field value $\delta_{\mathrm{s}} \big(\bm{x}, R(M_{\mathrm{h}})\big)$ exceeds the collapse barrier, $p \big[ \delta_{\mathrm{s}} \big(\bm{x}, R(M_{\mathrm{h}})\big)  > \delta_c(t) \big] $, equals to the fraction of mass materials contained in haloes with $M > M_{\mathrm{h}}$.
 The smoothed field is defined as
\begin{align}
    \delta_{\mathrm{s}} \big(\bm{x}, R(M_{\mathrm{h}})\big) \equiv \int \delta_{\mathrm{init}} (\bm{x}') \, W(\bm{x} - \bm{x}'; R \big(M_{\mathrm{h}}) \big) \rd^3 \bm{x}' \ ,
\end{align}
where $W(\bm{x}; R)$ is a filter (window function) with smoothing scale $R$ corresponding to halo mass  $M_{\mathrm{h}} = \gamma_{\mathrm{f}} \bar{\rho}_{\mathrm{m}} R^3$, with $\gamma_{\mathrm{f}} = 4\pi/3$ for top-hat filter and $\gamma_{\mathrm{f}} = 6\pi^2$ for sharp $k$-space filter.

If $\delta_{\mathrm{init}} (\bm{x})$ is a Gaussian random field then it is spectified by its (linear) power spectrum $P(k)$, and $\delta_{\mathrm{s}} (\bm{x})$ is also Gaussian according to its definition. 
The variance of the smoothed overdensity field $\sigma^2 (M_{\mathrm{h}})$ represents the typical fluctuation amplitude smoothed on scale $R \sim M_{\mathrm{h}}$, which is given by 
\begin{align}
    S \big( R(M_{\mathrm{h}}) \big) &\equiv \sigma^2 (R) \equiv \langle \delta^2_{\mathrm{s}} (\bm{x}; R) \rangle \notag  \\
    &= \frac{1}{2\pi^2} \int_0^\infty P(k) \tilde{W}^2 (kR) k^2 \rd k \ , \label{equ:sdfsdsd}
\end{align}
where $\tilde{W} (kR)$ is the Fourier transform of the $W(\bm{x}; R)$. 
If the linear power specturm $P(k)$ is given, $S, \sigma, R$ and $M_{\mathrm{h}}$ are equivalent measures of the smoothing scale and the assigned mass to haloes. They will be used interchangeably below.

The idea of \citet{1974ApJ...187..425P} suffers from a `fudge-factor' problem. The original Press-Schechter postulate predicts that only $1/2$ of all matter in the Universe is locked-up in collapsed haloes. They `solved' this problem by introducing a fudge factor two, i.e., relating the mass fraction with $2 \times p \big[ \delta_{\mathrm{s}} \big(\bm{x}, R(M_{\mathrm{h}})\big)  > \delta_c(t) \big]$. 
The excursion set formalism, came up with by \citet{1991ApJ...379..440B}, provides an alternative derivation of the halo mass function that truely solves the `fudge-factor' problem.

Without loss of generality, we will consider the halo mass function at present day hereafter, since the discussion below is valid for any time. We denote the initial overdensity field extrapolated to today as $\delta(\bm{x})$ and smoothed field as $\delta_{\mathrm{s}} (\bm{x}; S)$, following standard literatures. 
Considering a location $\bm{x}$, the smoothed overdensity $\delta_{\mathrm{s}} (\bm{x}; S)$ is a trajectory of random walk in $\delta_{\mathrm{s}} \text{-} S$ space.
In the limit $S \to 0$, which corresponds to $M \to \infty$ in hierarchical struction formation cosmologies such as $\Lambda$CDM, $\delta_{\mathrm{s}} (\bm{x}; S) \to 0$ for any $\bm{x}$. So the random walk of can be viewed as starting from $(S = 0, \delta_{\mathrm{s}} = 0)$, when increasing $S$ (corresponding to decreasing the halo mass), $\delta_{\mathrm{s}}$ wanders away from zero. 
A plot of the smoothed density versus the size of the filter $S(R)$ traces out a random walk.

In the spirit of Press-Schechter formalism, a spherical region of initial radius $R$ whose center located in $\bm{x}$ is considered to have collapsed to a virialized object today or live in a larger region which has collapsed earlier if $\delta_{\mathrm{s}} (\bm{x}; S(R)) > \delta_c$, where the collapse barrier $\delta_c$ is solved from spherical or ellipsoidal collapse discussed in last section. 
The ansatz of excursion set formalism is that, the fraction of trajectories with a \textit{first} crossing of the collapse barrier $\delta_c$ at $S > S_1 = \sigma^2 (M_1)$ is equal to the mass fraction of haloes with masses $M < M_1$. Denoting the mass fraction as $F(<M_1) = 1 - F(>M_1)$, the predicted halo mass function is \citep{2010gfe..book.....M} 
\begin{align}
    \frac{\rd n(M)}{\rd M} \rd M &= \frac{\bar{\rho}_{\mathrm{m0}}}{M} \frac{\partial F(>M)}{\partial M} \rd M = \frac{\bar{\rho}_{\mathrm{m0}}}{M} f(S, \delta_c) \rd S \ ,
\end{align}
where $f(S, \delta_c) \rd S$ is the the probability that the random walk $\delta_{\mathrm{s}}(S)$ first crosses the barrier at the interval $(S, S+\rd S)$. Note that the halo mass function $\frac{\rd n(M)}{\rd M}$ is denoted as $n(M)$ in some literatures, e.g. \citet{2001MNRAS.323....1S,2010gfe..book.....M}, which may cause confusion.

Given the collpase barrier $\delta_c$, the first-crossing probability $f(S)$ can be obtained by the Monte Carlo simulation, i.e., simulating many trajectories $\{ \delta_{\mathrm{s}} (\bm{x}_i; S)$ for $i=1, 2, \dots, N \}$. \citet{2006ApJ...641..641Z} derived an elegant formulation for $f(S)$ with arbitrary shape of collapse barrier $\delta_c = B(S)$:
\begin{align}
    f(S) &= g(S) + \int_0^S \rd S' f(S') h(S, S') \ , \\
\intertext{in which}
    g(S) &\equiv \left[ \frac{B(S)}{S} - 2\frac{\rd B}{\rd S} \right] P_0 \big[B(S), S \big] \ , \\
    h(S, S') &\equiv \left[ 2\frac{\rd B(S)}{\rd S} - \frac{B(S) - B(S')}{S - S'} \right] P_0 \big[B(S)-B(S'), S - S' \big] \ .
\end{align}
where 
\begin{align}
    P_0 (\delta, S) = \frac{1}{\sqrt{2\pi S}} \exp \left( - \frac{\delta^2}{2S} \right)
\end{align}
is the Gaussian distribution. In our following calculation, we will adopt \citet{2006ApJ...641..641Z} algorithm to compute the mass function.

\subsection{Sheth-Tormen formula as a good approximation for ellipsoidal collapse barrier in $f(R)$ gravity}
Collapse barriers are solved from gravitational collapse of top-hat overdensities. In the simplest case, spherical collapse in $\Lambda$CDM background, $\delta_{\mathrm{sc}}^{\Lambda\mathrm{CDM}} \approx 1.676$ is constant\footnote{In the Einstein-de Sitter universe, $\delta_{\mathrm{sc}}^{\mathrm{EdS}} \approx 1.686$.}. 
\citet{2001MNRAS.323....1S} suggested that the ellipsoidal collapse would substantially improve the predicted halo mass function compared with simulation. As described in \S~\ref{sec:GREC}, the ellipsoidal collapse barrier $\delta_{\mathrm{ec}}^{\mathrm{GR}} (e, p)$ depends on the surrounding shear field, which is characterized by ellipsoidal-geometry-related parameters $e$ and $p$ of the collpased region.
The full excursion set random walk should proceed in this high-dimensional parameter space. 
\citet{2001MNRAS.323....1S} considered the averaged collapse barrier by averaging $\delta_{\mathrm{ec}}^{\mathrm{GR}} (e, p)$ over the distribution of $e$ and $p$ of a Gaussian field. 
Gaussian field $\delta_{\mathrm{s}}$ smoothed on the scale $M_{\mathrm{h}}$ has variance $\sigma^2 (M_{\mathrm{h}}; a_{\mathrm{init}})$. In this field, regions initially having a given overdensity $\delta_{\mathrm{init}} / \sigma (M_{\mathrm{h}}; a_{\mathrm{init}})$ have a most probable ellipticity $e_{\mathrm{mp}} = \sigma (M_{\mathrm{h}}; a_{\mathrm{init}}) / (\sqrt{5} \delta_{\mathrm{init}})$ and $p_{\mathrm{mp}} = 0$ (see Appendix~A of \citet{2001MNRAS.323....1S}), i.e.,
\begin{align}
    e_{\mathrm{mp}} &= \frac{\sigma(M_{\mathrm{h}}; a_{\mathrm{init}})}{\sqrt{5} \delta_{\mathrm{init}}} = \frac{\sigma(M_{\mathrm{h}}; a_0)}{\sqrt{5} \delta_{\mathrm{ec}}^{\mathrm{GR}}} \ , \\ 
    \intertext{and} 
    p_{\mathrm{mp}} &= 0 \ ,
\end{align}
respectively. 
To relate $e$ and $p$ to the mass $M_{\mathrm{h}}$ or $S$, \citet{2001MNRAS.323....1S} replaced $e$ and $p$ with their most probable values in Equation~\eqref{equ:STori}, which yields 
\begin{align}
    \delta_{\mathrm{ec}}^{\mathrm{GR}} (e,p) &\xrightarrow[p\sim p_{\mathrm{mp}}=0]{S \equiv \sigma^2(M_{\mathrm{h}}) \sim (\sqrt{5} e \delta_{\mathrm{ec}}^{\mathrm{GR}})^2} \delta_{\mathrm{ec}}^{\mathrm{GR}} (S)   \ , \\
    \delta_{\mathrm{ec}}^{\mathrm{GR}} (S) &= \delta_{\mathrm{sc}}^{\mathrm{GR}} \left(1 + \beta \left[ \frac{S}{\delta_{\mathrm{sc}}^{\mathrm{GR}}} \right]^{\gamma} \right) \ . \label{equ:ST1}
\end{align}
This deviation caused by neglecting scatter around the most probable value has been tested, which shows that Equation~\eqref{equ:ST1} is a rather good approximation \citep{2002MNRAS.329...61S}.
Under this replacement the high-dimensional random walk is evaded and we can still use the method of \citet{2006ApJ...641..641Z} to calculate the first-crossing distribution.

In order to improve the consistency between the prediction of  the excursion set theory and $N$-body simulation, \citet{2001MNRAS.323....1S} found that it is necessary to introduce a new parameter $a \approx 0.707$ ($a \approx 0.75$ in \citet{2009PhRvD..79h3518S}), and postulate the form of the collapse barrier is rather
\begin{align}
    \delta_{\mathrm{ec}}^{\mathrm{GR}} (S) = \sqrt{a} \delta_{\mathrm{sc}}^{\mathrm{GR}} \left( 1 + \beta \left[ \frac{S}{\sqrt{a} \delta_{\mathrm{sc}}^{\mathrm{GR}}} \right]^{\gamma}  \right) \ . \label{equ:st13}
\end{align}
The parameter $a$ is not derived from the ellipsoidal collapse but introduced by hand in order to fit the $N$-body simulation results. \citet{2010ApJ...717..515M} argued that the parameter $a$ can be explained by considering the collapse barrier itself as a stochastic variable. 

The Sheth-Tormen formula Equation~\eqref{equ:ST1} encodes the ingredients of ellipsoidal collapse in GR as a function of the spherical collapse barrier $\delta_{\mathrm{sc}}$. 
When extending the excursion set formalism from GR to MG, we usually assume that the effects of ellipsoidal collapse and MG on the collapse barrier can be treated separately \citep{2013PhRvD..87l3511L,2014JCAP...04..029B,2014JCAP...03..021L,2018MNRAS.476L..65H}. 
That is, replacing $\delta_{\mathrm{sc}}^{\mathrm{GR}}$ in Sheth-Tormen formula with the spherical collapse barrier in MG. For example, we assume that the ellipsoidal collapse barrier in $f(R)$ gravity is given by
\begin{align}
    \delta_{c}^{\mathrm{ST}} (S) \equiv \sqrt{a} \delta_{\mathrm{sc}}^{{f(R)}} (S) \left( 1 + \beta \left[ \frac{S}{\sqrt{a} \delta_{\mathrm{sc}}^{f(R)}(S) } \right]^{\gamma}  \right) \ . \label{equ:st2}
\end{align}
The spherical collapse barrier $\delta_{\mathrm{sc}}^{f(R)}$ in $f(R)$ gravity is a function of halo mass $M_{\mathrm{h}}$ and environmental overdensity $\delta_{\mathrm{env}}$, due to the existance of the environment-dependent fifth force. 

To check the validity of Sheth-Tormen formula \eqref{equ:st2} as an approximation of ellipsoidal collapse in $f(R)$ gravity, we directly solve the ellipsoidal collapse process and inspect the behavior of the `true' collapse barrier $\delta_{\mathrm{ec}}^{f(R)}$.
In \S~\ref{sec:FREC} we present a simple modeling of this, and the corresponding critical value is $\delta_{\mathrm{ec}}^{f(R)} = \delta_{\mathrm{ec}}^{f(R)} (M_{\mathrm{h}}, \delta_{\mathrm{env}}, e, p)$, which combines the ingredients of ellipsoidal collapse and modified gravity. 
In the same spirit of \citet{2001MNRAS.323....1S}, we would recast these variables into one variable $S$ by relating their most probable values. 
First, $e$ and $p$ can be approximated by their most probable value as in Equation~\eqref{equ:ST1}. 

Second, the environment overdensity is related to the definition of the radius of environment. 
We adopt the definition used in \citet{2012MNRAS.425..730L}, that is, defining the radius by environment's Eulerian (physical) radius $\zeta = 5 \, h^{-1}\,\mathrm{Mpc}^{-1}$ at $z = 0$.  
The probability distribution of $\delta_{\mathrm{env}}$ and its approximate analytical expressions can be found in \citep{2008MNRAS.386..407L,2012MNRAS.426.3260L,2012MNRAS.425..730L}. 
Assuming cosmological parameter values as defined in Section~\ref{sec:FREC}, the most probable value $\delta_{\mathrm{env, mp}} \approx 0.8$ \citep{2014JCAP...03..021L}. We adopt this most probable value as an approximation of $\delta_{\mathrm{env}}$. 


Now, we are in the position that the collapse barrier is a function of both variance $S$ and halo mass $M_{\mathrm{h}}$,  $\delta_{\mathrm{ec}}^{f(R)} = \delta_{\mathrm{ec}}^{f(R)}  (S, M_{\mathrm{h}})$.
Note that $S$ and $M_{\mathrm{h}}$ is related via the intergation of linear power spectrum Equation~\eqref{equ:sdfsdsd}. 
Thus we can recast $S$ and $M_{\mathrm{h}}$ into one variable $S$:
\begin{align}
    \delta_{\mathrm{ec}}^{f(R)} (M_{\mathrm{h}}, \delta_{\mathrm{env}}, e, p) &\xrightarrow[p \sim p_{\mathrm{mp}}=0]{\sigma(M) \sim \sqrt{5} e \delta_{\mathrm{ec}}^{f(R) } } \delta_{\mathrm{ec}}^{f(R)} (M_{\mathrm{h}}, \delta_{\mathrm{env}}, S) \notag \\
    & \xrightarrow{\delta_{\mathrm{env}} \sim \delta_{\mathrm{env, mp}}} \delta_{\mathrm{ec}}^{f(R)} (M_{\mathrm{h}}, S) \ , \notag \\
    & \xrightarrow{S = S(M_{\mathrm{h}})} \delta_{\mathrm{ec}}^{f(R)} (S) \ .
\end{align}

\begin{figure}
\centering 
\includegraphics[width=\columnwidth]{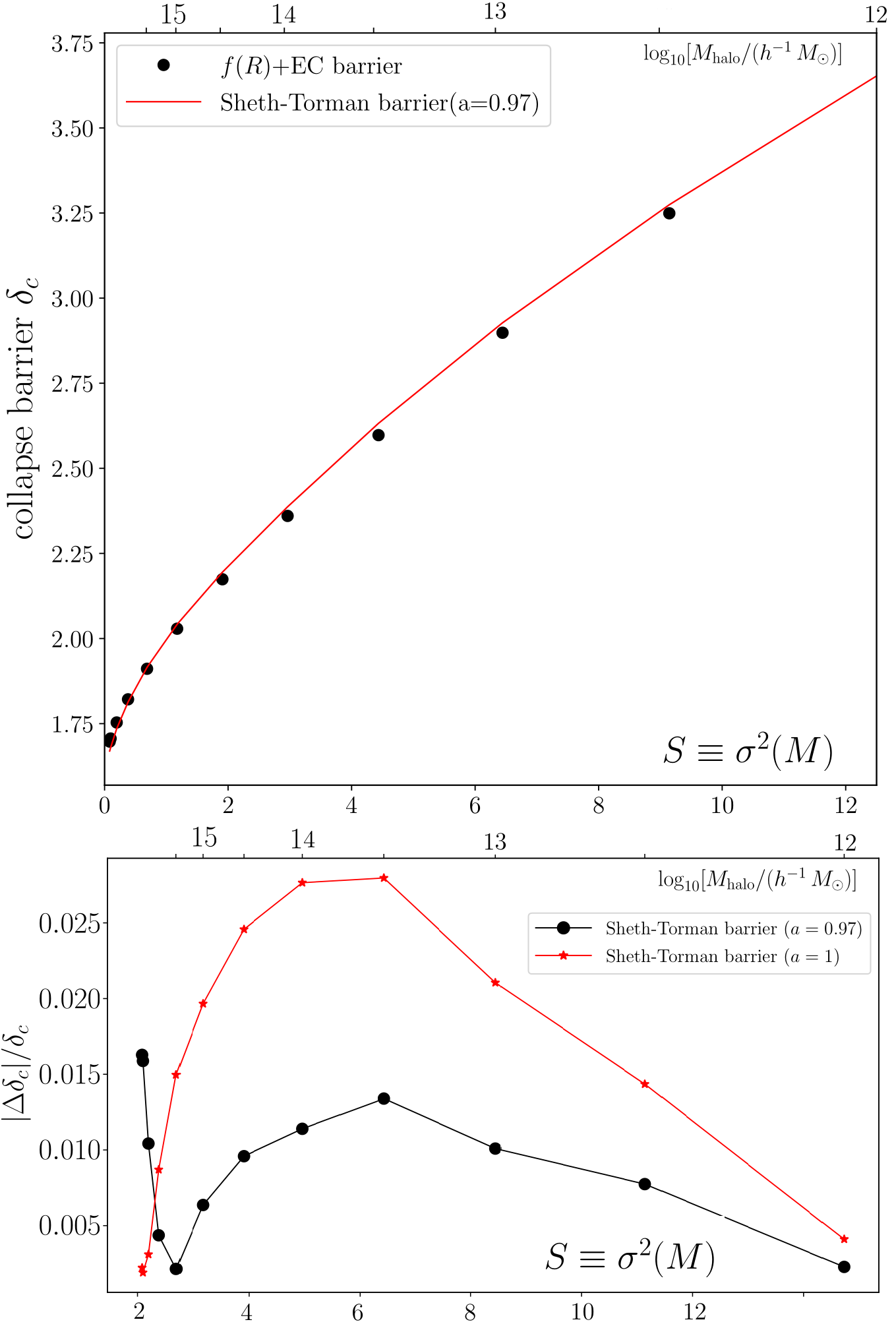}
\caption{ Comparison of collpase barriers from Sheth-Tormen (ST) formula and full $f(R)$ gravity Ellipsoidal Collapse (EC) dynamics modeling. The small relative differences ($
\Delta \delta_c \equiv |\delta_c^{\mathrm{ST}} - \delta_{\mathrm{ec}}^{f(R)}|/\delta_{\mathrm{ec}}^{f(R)} \lesssim 2.7\%$ for $a=1$ and $\lesssim 1.6\%$ for $a=0.97$ in ST formula) indicate that the Sheth-Tormen formula is a good approximation of ellipsoidal collapse in $f(R)$ gravity.}
\label{fig:deltac_and_reldiff}
\end{figure}

We find a good consistency between the `true' ellipsoidal collapse barrier $\delta_{\mathrm{ec}}^{f(R)} (S)$ (recast by replacing other variables with their most probable values) and the Sheth-Tormen approximation Equation~\eqref{equ:st2}. Figure~\ref{fig:deltac_and_reldiff} shows the comparison of two barriers and their relative difference $|\Delta \delta_c|/\delta_c \equiv |\delta_{\mathrm{ec}}^{f(R)} - \delta_c^{\mathrm{ST}}| / \delta_{\mathrm{ec}}^{f(R)}$, with $|f_{R0}| = 10^{-5}$. When the parameter $a$ in Sheth-Tormen equation is equal to one, the largest deviation is $2.7\%$. And the optimized value of $a$ is $0.97$, which corresponds to $\lesssim 1.6\%$ relative difference. 
Note that the Sheth-Tormen formula truely used in the excursion set formalism needs a calibriated value of parameter~$a$ such as $0.75$, which can not be described by the simple ellipsoidal collapse of top-hat overdensity.
We simply set $a = 0.75$ following \citet{2009PhRvD..79h3518S} when calculating the first-crossing probability.
Our results show that Sheth-Tormen formula is a good approximation for ellipsoidal collapse barrier in $f(R)$ gravity. 
This means that one does not need to cope with the complex ellipsoidal collapse, at least in $f(R)$ gravity. This is the main conclusion of this paper.

\subsection{Non-dynamical approximation}
\label{subsec:nondynapprox}
The Sheth-Tormen formula, summarizing the scale dependence of the ellipsoidal collapse critical value, still needs spherical collapse barrier in $f(R)$ gravity.
The spherical collapse barrier $\delta_{\mathrm{sc}}^{f(R)}$, defined in Equation~\eqref{equ:deltascfr}, varies in the range of $(1.676, 1.692)$, with only $\sim 1\%$ relative amplitude \citep{2018MNRAS.476L..65H}. 
We argue that this small variation is cause by the offset of two effects caused by gravity enhancement in $f(R)$ gravity. 

First, when solving the spherical collapse equations of top-hat overdensities, one shall adjust the initial overdensity $\delta_{\mathrm{h,init}}$ so that makes the halo collapse at $z=0$. 
Under the same conditions, $\delta_{\mathrm{init}}$ in $f(R)$ gravity is smaller than that in GR, since the gravity in $f(R)$ gravity is stronger and the gravitational collapse is faster.
Second, for the same reason, the linear growth function $D (a, k)$ of $f(R)$ gravity is larger than the $\Lambda$CDM case. Recall that the collapse barrier $\delta_{\mathrm{sc}}^{f(R)} \equiv D(a_0, k_{\mathrm{h}};a_{\mathrm{init}}) \delta_{\mathrm{h,init}}$. 
Thus, in $f(R)$ gravity, the greater $D(a, k)$ and smaller $\delta_{\mathrm{h, init}}$ cancel out each-other. 

\begin{figure}
    \centering 
    \includegraphics[width=\columnwidth]{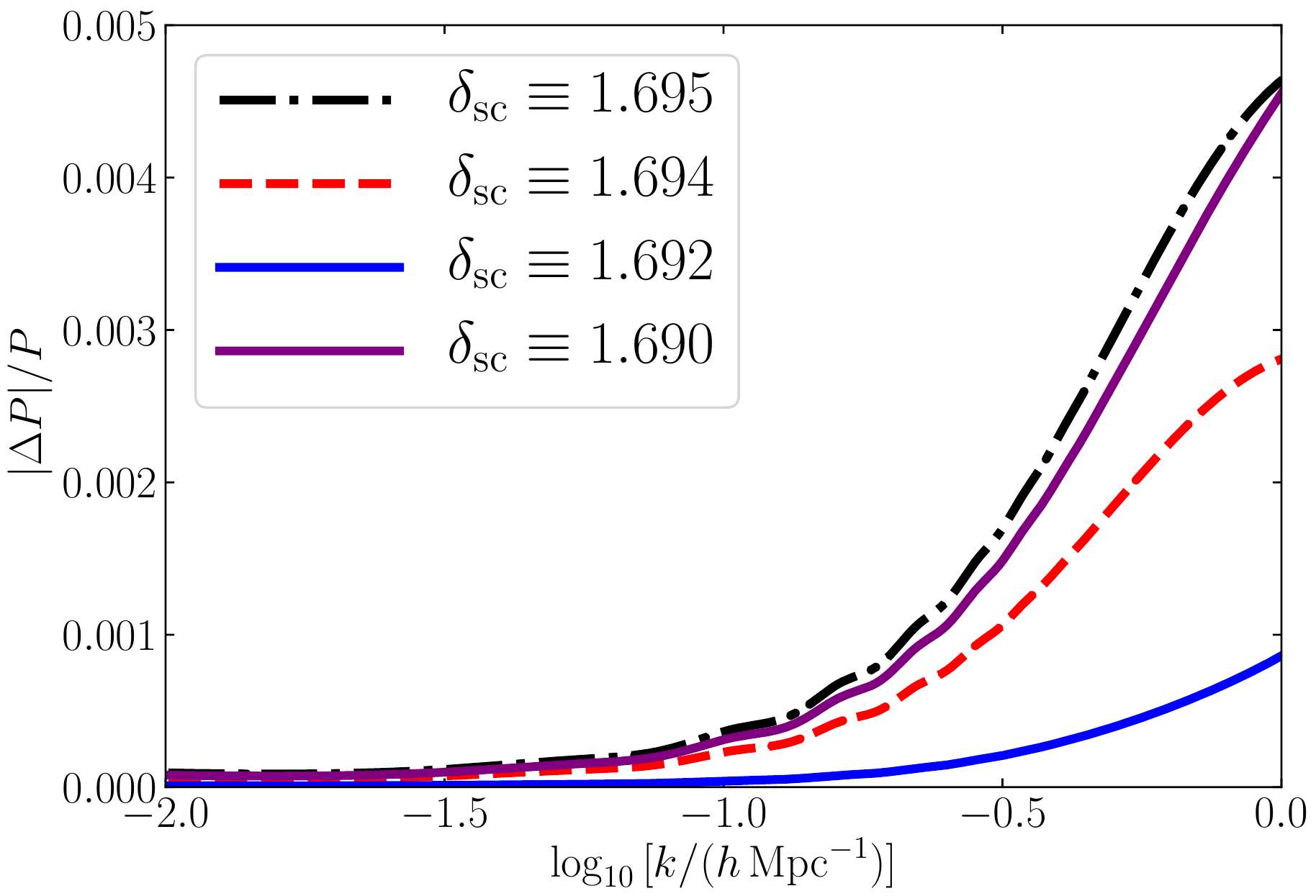}
    \caption{ The relative difference of non-linear power spectra in $f(R)$ gravity with $|f_{R0}| = 10^{-4}$ (F4) calculated by CHAM between full scenario and non-dynamical approximation, $|\Delta P| / P$, where $\Delta P \equiv P^{\mathrm{full}} - P^{\mathrm{non} \text{-} \mathrm{dyn}}$. The non-dynamical approximation, i.e., a constant $\delta^{f(R)}_{\mathrm{sc}}$ instead of a function of halo mass, causes $0.5\%$ relative deviation at most.}
    \label{fig:nondyn_diffdeltaSC_F4}
\end{figure}

\begin{figure}
        \centering 
        \includegraphics[width=\columnwidth]{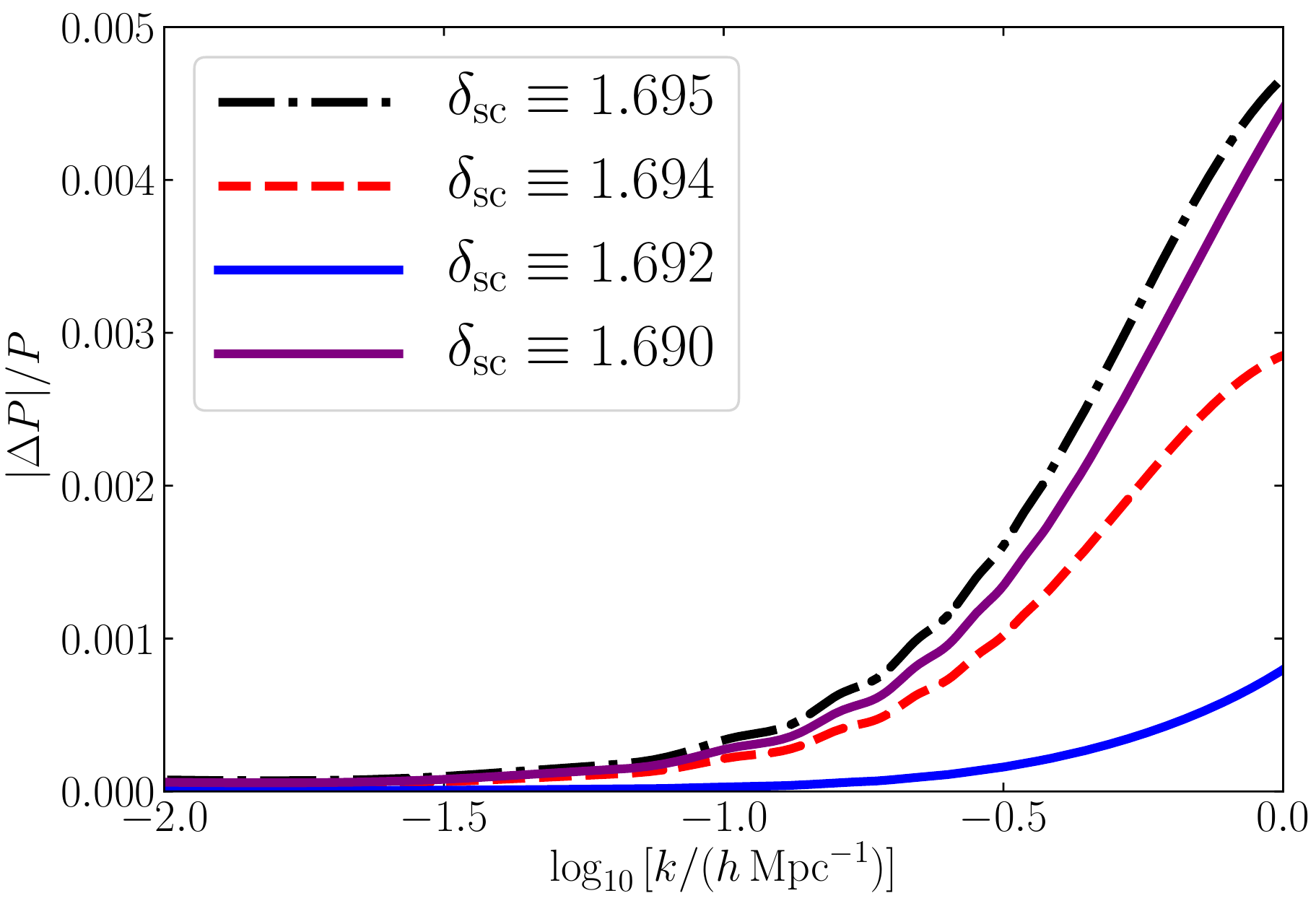}
        \caption{ Same as Figure~\ref{fig:nondyn_diffdeltaSC_F4} but for $f(R)$ gravity with $|f_{R0}| = 10^{-5}$.}
        \label{fig:nondyn_diffdeltaSC_F5}
\end{figure}

Since $\delta_{\mathrm{sc}}^{f(R)} (M_{\mathrm{h}})$ is insensitive to mass or scale, we adopt a \textit{non-dynamical approximation}, in which $\delta_{\mathrm{sc}}^{f(R)}$ is approximated by a constant. 
The rest of the calculation, such as the mass function, linear bias and concentration, are the same as \cite{2018MNRAS.476L..65H}. 
We have checked that this approximation causes little change on non-linear matter power spectrum from the original full scenario presented by \citet{2018MNRAS.476L..65H}. 
The linear power spectrum is output from the \texttt{EFTCAMB}\footnote{\url{http://eftcamb.org/}} Hu-Sawicki $f(R)$ module \citep{Hu:2016zrh}.
As shown in Figure~\ref{fig:nondyn_diffdeltaSC_F4} and \ref{fig:nondyn_diffdeltaSC_F5}, the relative differences $|\Delta P| / P$ is less than $0.1\%$ up to $k = 1\, [h/{\rm Mpc}]$, when the optimized value $\delta_{\mathrm{sc}}^{f(R)} \approx 1.692$ is adopted, which is the exact value predicted in \citet{2018MNRAS.476L..65H}. This is another main conclusion of the paper.

\subsection{Comparison with N-body simulation results}
\begin{figure}
    \centering 
    \includegraphics[width=\columnwidth]{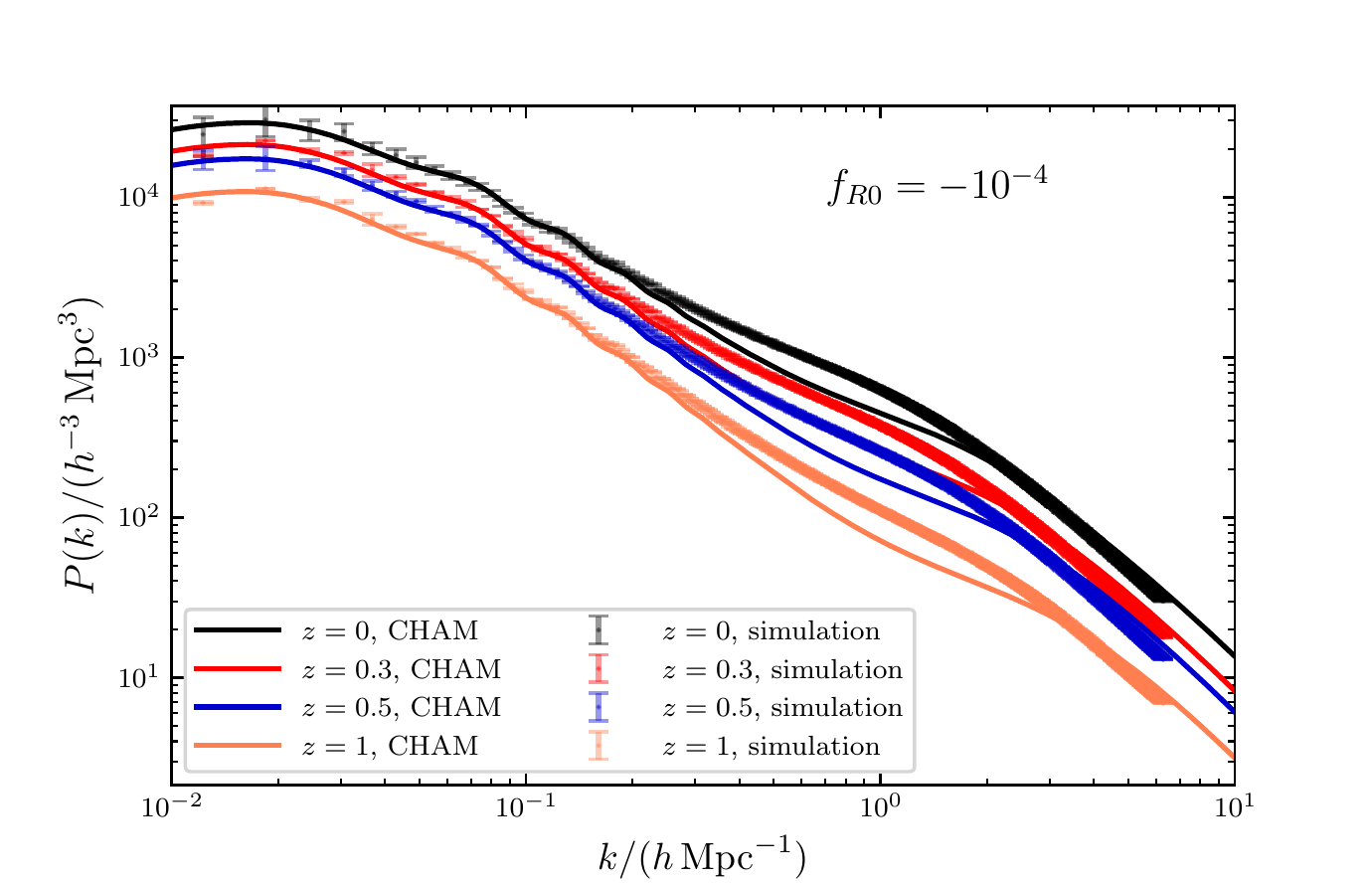}
    \caption{ Matter spectrum comparison for $f(R)$ gravity with $f_{R0} = -10^{-4}$ (F4). 
    Data points are from \texttt{ELEPHANT} simulation, and curves are outputs of \texttt{CHAM}. }
    \label{fig:pok_F4}
\end{figure}

\begin{figure}
    \centering 
    \includegraphics[width=\columnwidth]{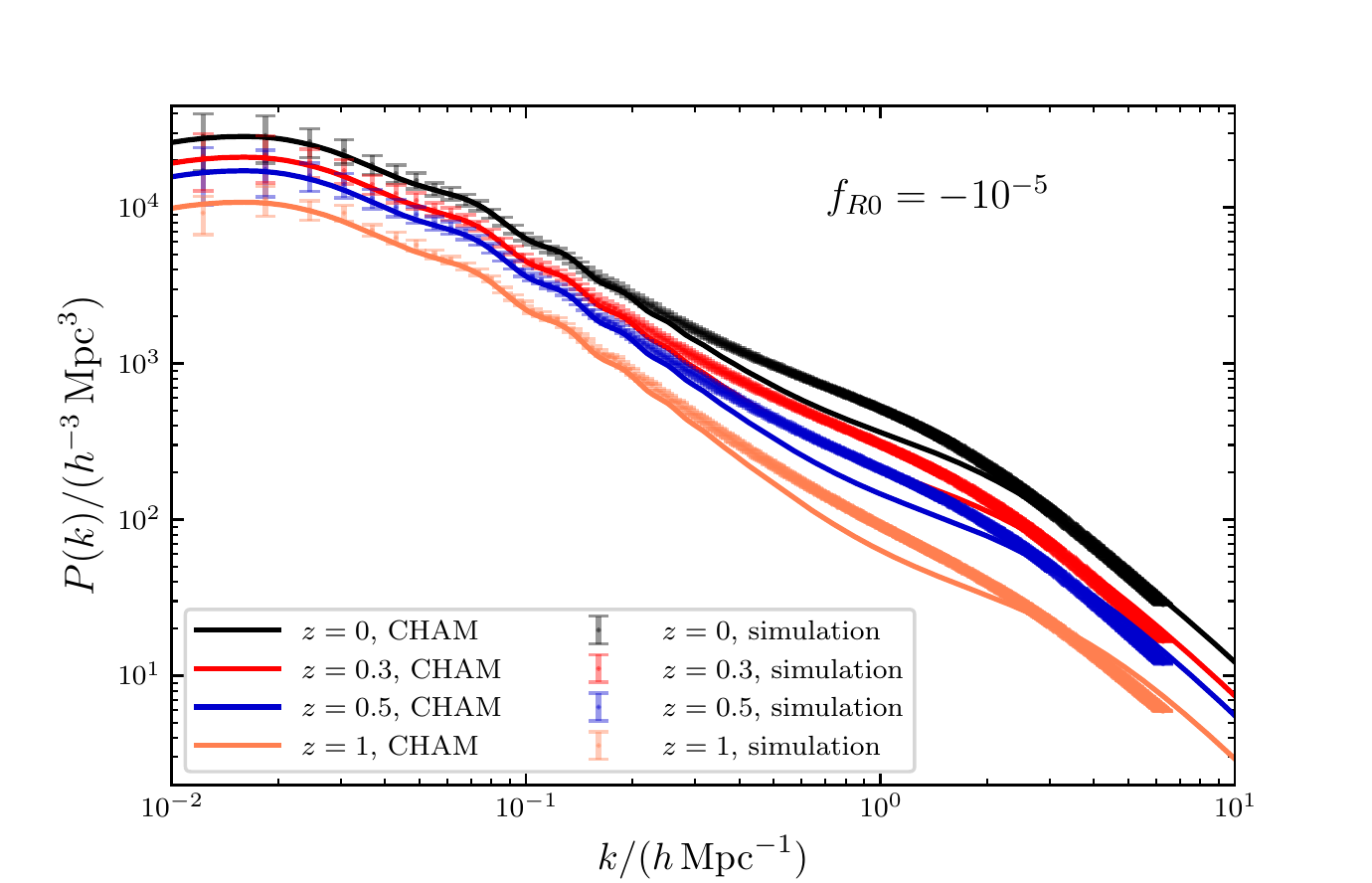}
    \caption{ Matter spectrum comparison for $f(R)$ gravity with $f_{R0} = -10^{-5}$ (F5). 
    Data points are from \texttt{ELEPHANT} simulation, and curves are outputs of \texttt{CHAM}. }
    \label{fig:pok_F5}
\end{figure}

\begin{figure}
    \centering 
    \includegraphics[width=\columnwidth]{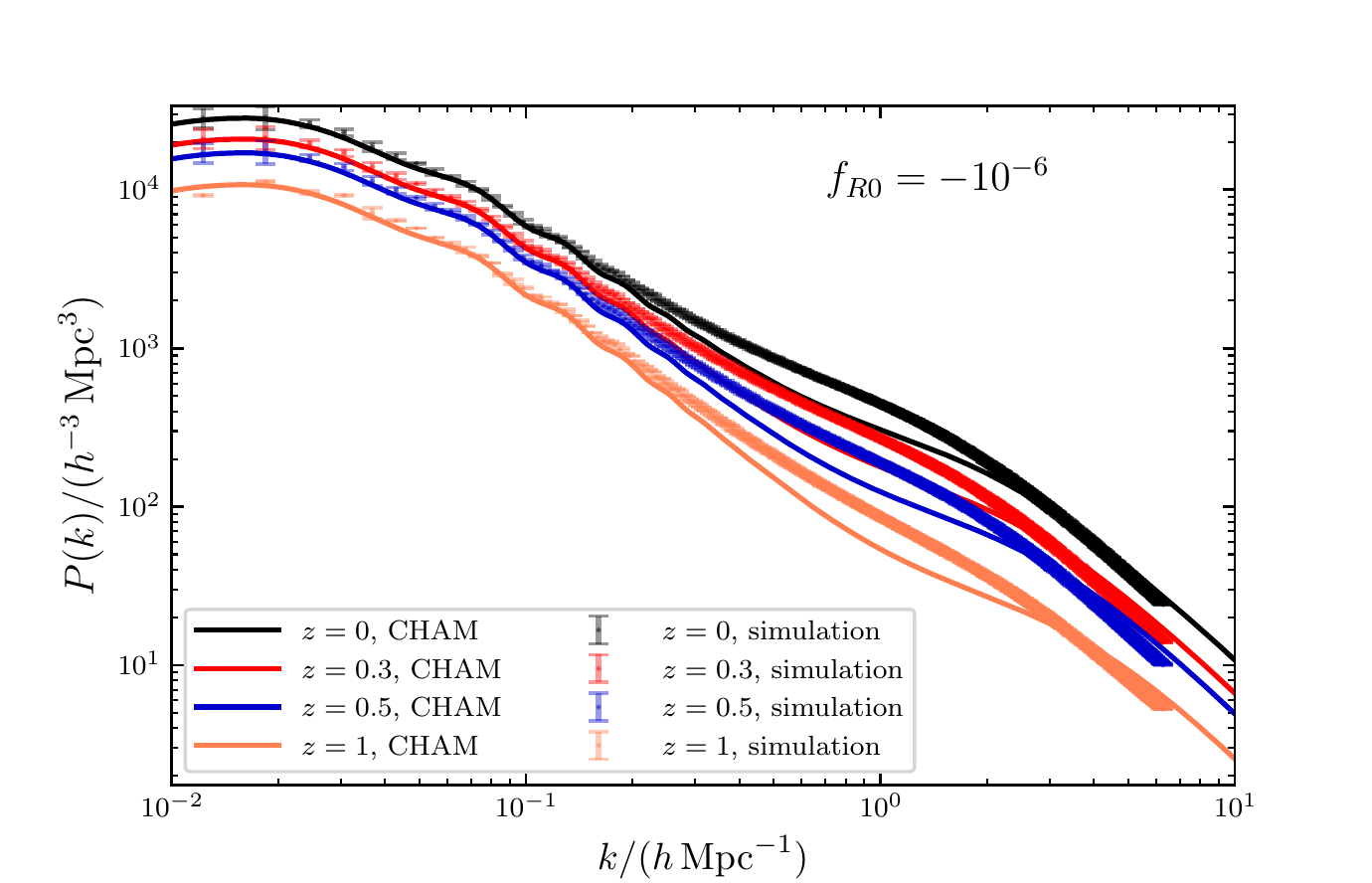}
    \caption{ Matter spectrum comparison for $f(R)$ gravity with $f_{R0} = -10^{-6}$ (F6). 
    Data points are from \texttt{ELEPHANT} simulation, and curves are outputs of \texttt{CHAM}. }
    \label{fig:pok_F6}
\end{figure}

\begin{figure}
    \centering 
    \includegraphics[width=\columnwidth]{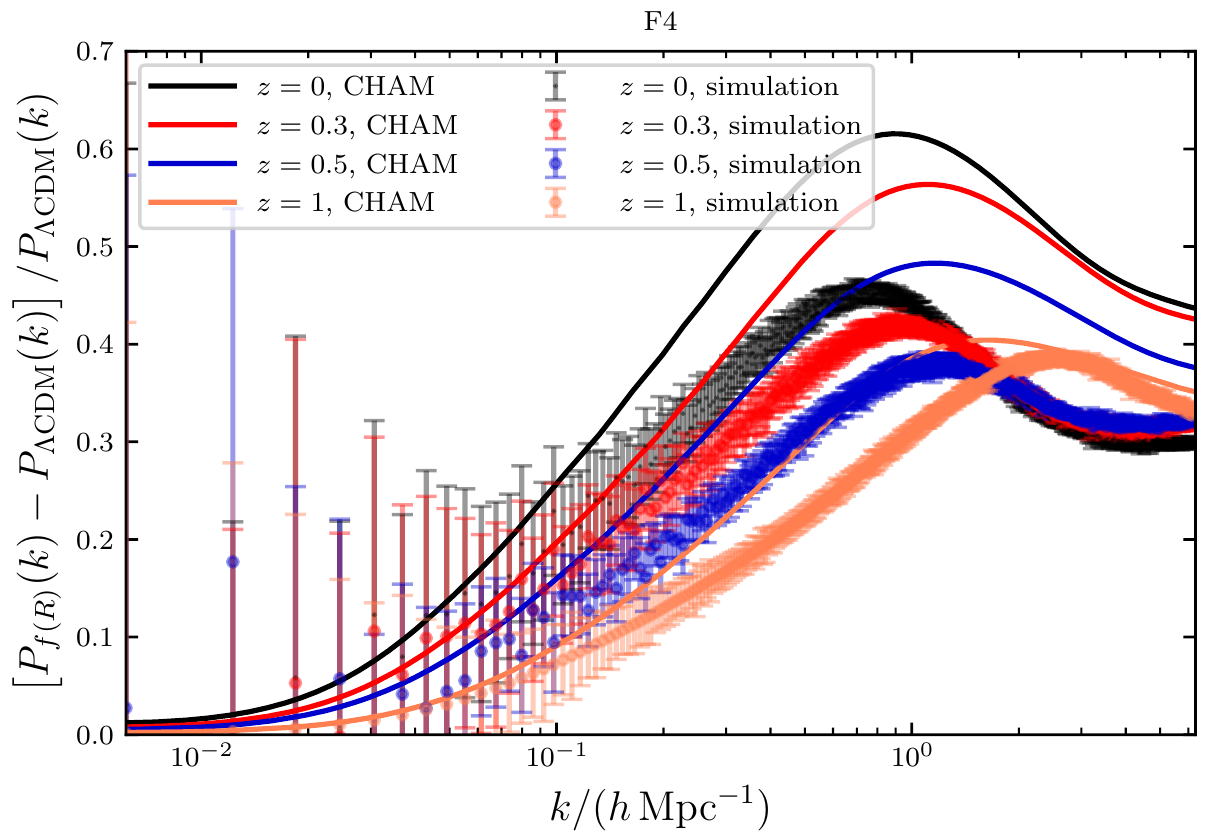}
    \caption{Relative matter spectrum comparison for $f(R)$ gravity with $f_{R0} = -10^{-4}$ (F4). 
    Data points are from \texttt{ELEPHANT} simulation, and curves are outputs of \texttt{CHAM}. }
    \label{fig:Dpok_F4}
\end{figure}

\begin{figure}
    \centering 
    \includegraphics[width=\columnwidth]{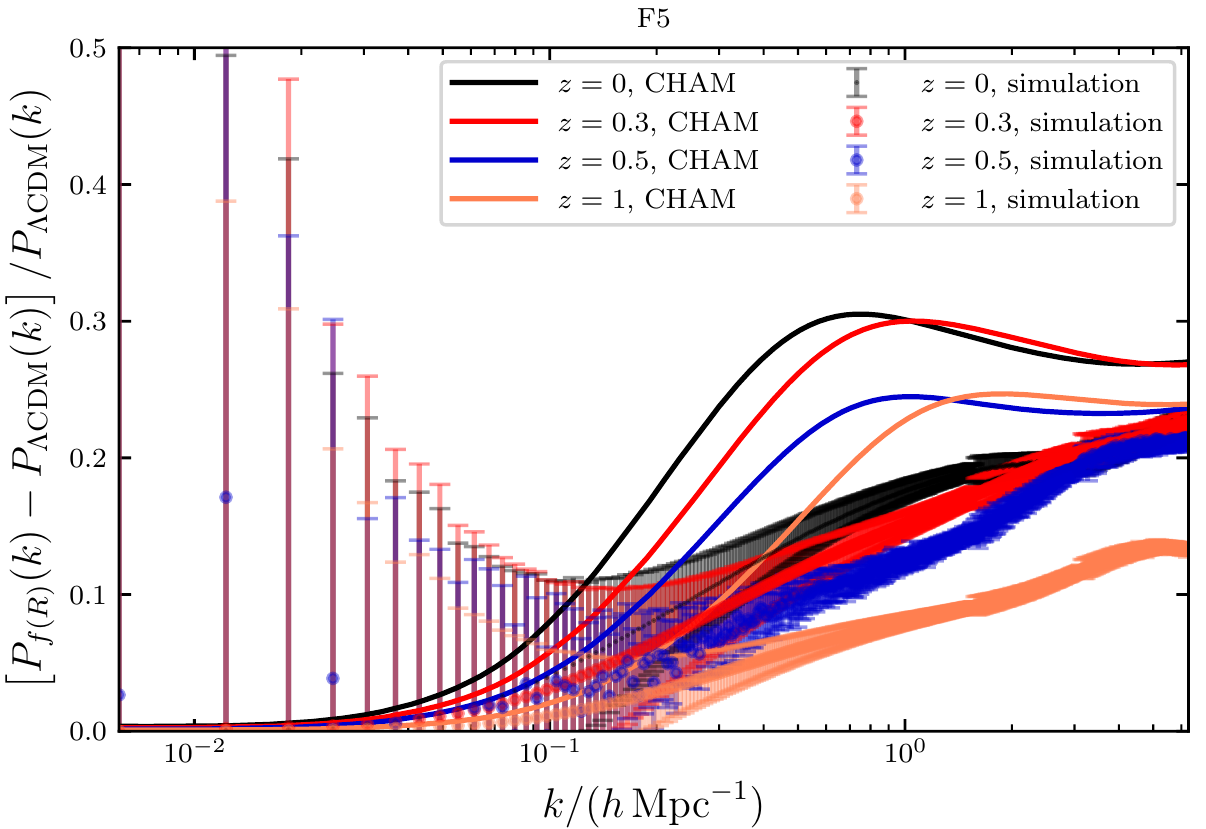}
    \caption{Relative matter spectrum comparison for $f(R)$ gravity with $f_{R0} = -10^{-5}$ (F5). 
    Data points are from \texttt{ELEPHANT} simulation, and curves are outputs of \texttt{CHAM}. }
    \label{fig:Dpok_F5}
\end{figure}

\begin{figure}
    \centering 
    \includegraphics[width=\columnwidth]{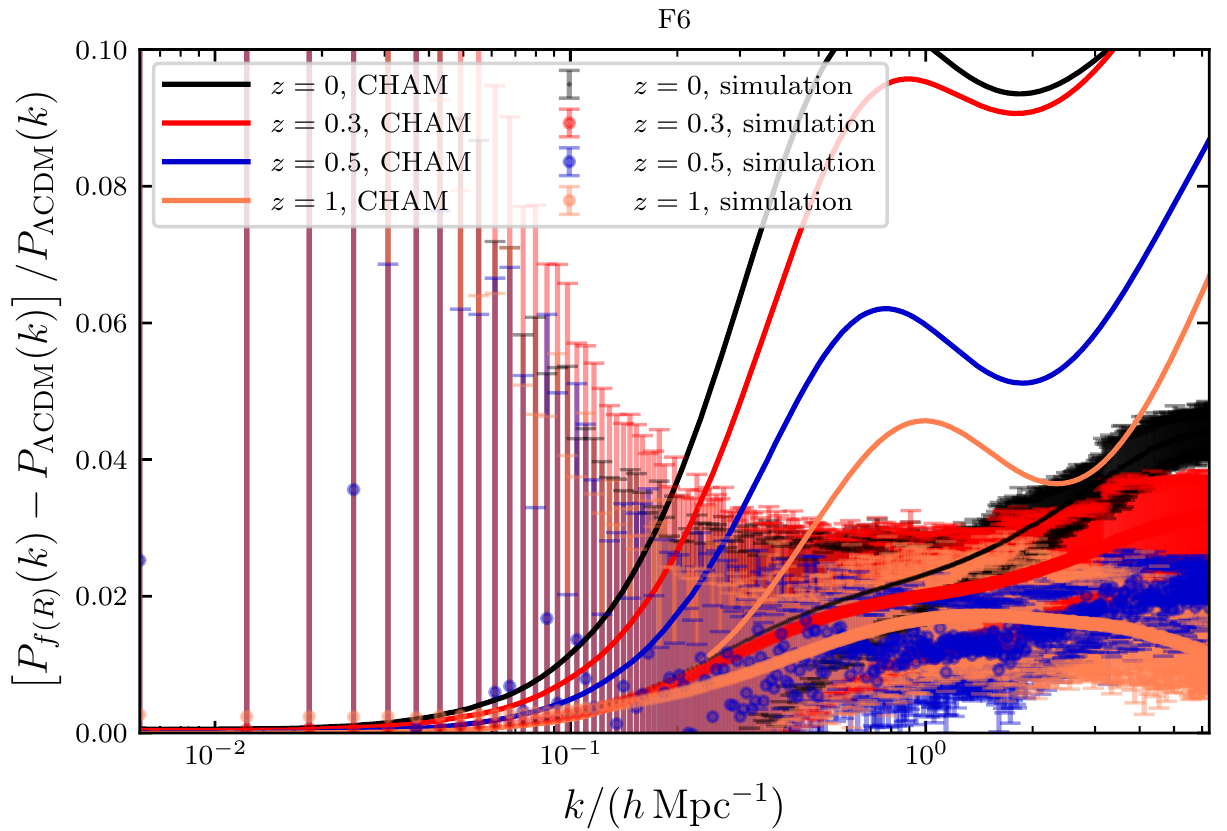}
    \caption{Relative matter spectrum comparison for $f(R)$ gravity with $f_{R0} = -10^{-6}$ (F6). 
    Data points are from \texttt{ELEPHANT} simulation, and curves are outputs of \texttt{CHAM}. }
    \label{fig:Dpok_F6}
\end{figure}

We use the Extended LEnsing PHysics using ANalaytic ray Tracing (\texttt{ELEPHANT}) dark matter only N-body simulations which have been run using the \texttt{ECOSMOG} \citep{2012JCAP...01..051L} and \texttt{ECOSMOG-V} \citep{2013JCAP...05..023L} codes for $f(R)$ gravity models. 
\texttt{ECOSMOG} and \texttt{ECOSMOG-V} are based on the adaptive mesh refinement N-body code \texttt{RAMSES} \citep{2002A&A...385..337T}. 
These codes are efficiently optimized and implemented with methods that speed up the calculations of the non-linear partial differential equations that characterize these models.
The cosmological parameters were adpoted from the WMAP9 year CMB measurements \citep{2013ApJS..208...19H}.
The simulations follow the evolution of $N_p = 1024^3$ particles with mass $m = 7.798\times10^{10} h^{-1} M_{\odot}$ in a cubical box of comoving size $L_{\rm box} = 1024 \,h^{-1} \mathrm{Mpc}$ from their initial conditions (generated with
the \texttt{MPGRAFIC} code, \citet{2008ApJS..178..179P}) at $z_{\rm ini}  = 49$ up to today ($z=0$).
Here, We compare the matter power spectrum outputs of \texttt{ELEPHANT} simulation and \texttt{CHAM}, at $z=0, 0.3, 0.5$ and $1$.
The full matter spectra are shown in Figures \ref{fig:pok_F4}, \ref{fig:pok_F5} and \ref{fig:pok_F6}, which correspond to $f_{R0} = -10^{-4}, -10^{-5}$ and $-10^{-6}$. 
We also highlight the spectrum relative differences computed from  \texttt{CHAM} and \texttt{ELEPHANT} simulation in Figures \ref{fig:Dpok_F4}, \ref{fig:Dpok_F5} and \ref{fig:Dpok_F6}. 

\begin{figure}
    \centering 
    \includegraphics[width=\columnwidth]{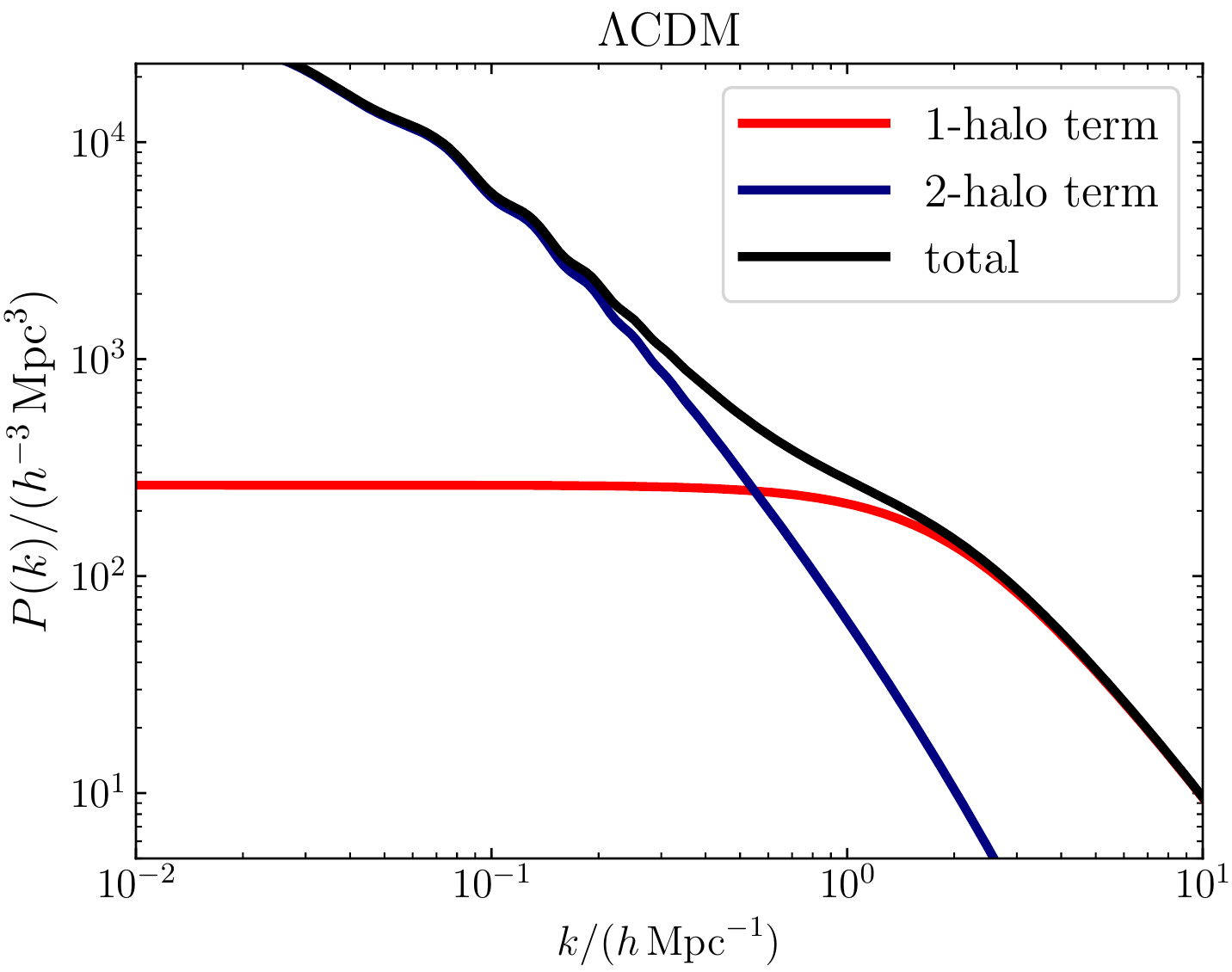}
    \caption{Demonstration of one halo term and two halo term.}
    \label{fig:haloterms}
\end{figure}

Via the halo model approach (Figures \ref{fig:pok_F4}, \ref{fig:pok_F5} and \ref{fig:pok_F6}), there exist a systematic under-estimation 
of the power spectrum in the co-moving wavenumber range between $0.3~h/{\rm Mpc}$ and $3~h/{\rm Mpc}$. 
From low to high redshifts, this discrepancy ends up in the larger wavenumber. 
Compared with our demonstration Figure \ref{fig:haloterms}, we can see that, this discrepancy regime is overlapped with
the transition scale between the two halo term and one halo term. 
We argue that this is due to the fact that our modified halo model is based on the original recipes~\cite{Sheth:1999mn,Bullock:1999he}. 
As shown in \cite{Mead:2015yca}, even in the $\Lambda$CDM framework, these problematic behaviors have already existed. 

There are several possible reasons for these poor performance. First of all, in simulations, the halo-finders normally only assign half of the particles
into haloes \cite{Jenkins:2000bv,More:2011dc}. Hence, the mass distribution of the other half of the N-body particles are treated via an extrapolated formula in the halo model. 
Secondly, by definition, the unvirialized objects are not taken into account in the halo model. This is the intrinsic drawback of this method. And these objects are expected to give essential 
contributions to the power spectrum in the mild non-linear regime. Besides of the cosmological parameters, the resulting spectra also relie on some astrophysical parameters, such as the halo 
concentration, \emph{etc}. How the measured power spectra are affected by these astrophysical parameter uncertainties and astrophysical assumptions have been investigated in \cite{Cooray:2000ry,Giocoli:2010dm,vanDaalen:2015msa,Pace:2015nda}.  

From Figure \ref{fig:Dpok_F4}, \ref{fig:Dpok_F5} and \ref{fig:Dpok_F6}, we can see that, in F4 model, N-body simulations give roughly $40\%$ relative differences w.r.t. $\Lambda$CDM. Whilst the \texttt{CHAM} predicts $60\%\sim40\%$ fractional differences
from low to high redshifts. In F5 model, N-body simulations give roughly $20\%$ differences at redshift below $0.5$ and $10\%$ difference at redshift $1$. Whilst the \texttt{CHAM} predicts $30\%$ differences below redshift $0.3$ and $20\%$ at redshifts $0.5$ and $1$.
In F6 model, the numbers in N-body simulations are below $5\%$ at all redshifts. Whilst the \texttt{CHAM} predicts $10\%\sim 5\%$ differences from low to high redshifts.

\section{Summary and Discussion}
\label{sec:summary}

In the previous work \citep{2018MNRAS.476L..65H}, we developed a fast numerical halo model algorithm (\texttt{CHAM}, which stands for the sCreened HAlo Model) for modeling non-linear matter power spectra for modified gravity cosmological models. 
In this paper, we examed one of the essential assumptions of \texttt{CHAM} --- using the Sheth-Tormen formula approximate the ellipsoidal collapse barrier in $f(R)$ gravity.
We model the ellipsoidal collapse of top-hat dark matter haloes in $f(R)$ gravity and calculate the more realistic collapse barrier. 
We find a good agreement between Sheth-Tormen formula and the `true' ellipsoidal collapse critical value in $f(R)$ theory. 
The relative difference of the ellipsoidal collapse barrier is less than or equal to $1.6\%$. 

Furthermore, we adopted the Sheth-Tormen collapse barrier formula and treated $\delta_{\mathrm{sc}}^{f(R)}$ as constant in halo mass. It means that we do not need to model the complicated ellipsoidal collapse process in $f(R)$ gravity. And all the modified 
gravity effect can be absorbed into the value shift in $\delta_{\mathrm{sc}}^{f(R)}$ compared with $\delta_{\mathrm{sc}}^{\rm GR}$. The calculation of the barrier shift is quite simple. We only need to rescale Newton constant by a factor $4/3$ in the linear matter 
density equation. Namely, treat the $\mu$ function in Equation (\ref{equ:Ddiff}) as a constant value. We call this assumption as `non-dynamical' version. The resulting non-linear spectra in F4 and F5 models, agree with the original full dynamical version of 
\texttt{CHAM} \citep{2018MNRAS.476L..65H} within $0.1\%$ precision up to $k=1[h/{\rm Mpc}]$. Due to the simplification of the non-linear dynamics modelling, the computational time of the code reduces significantly, from $10$ minutes to $1$ second.  
The updated version of the code can be found at \url{https://github.com/hubinitp/CHAM}. 
Finally, we compare our halo model prediction with N-body simulation. We find that the general spectrum profile agree, qualitatively. However, via the halo model approach, 
there exists a systematic under-estimation of the matter power spectrum in the co-moving wavenumber range between $0.3 h/{\rm Mpc}$ and $3 h/{\rm Mpc}$.
These scales are overlapping with the transition scales from two halo term dominated regimes to those of one halo term dominated. We argue that these mismatches are the discrepancies inherited from the original halo model. 
We will leave this problem for the future studies.
We think halo model is a physical intuitive approach and can help us understand the non-linear clustering process in the alternative theories to standard model. In the future, 
we plan to validate this method with more concrete models of modified gravity and dark energy.

\section*{acknowledgments}
We thank Baojiu Li for providing the simulation data. We also thank Pierluigi Monaco and Jie Wang for helpful discussions.
BH are supported by the Beijing Normal University Grant under the reference No. 312232102 and by the National Natural Science Foundation of China Grants No. 11973016, No. 11690023 and No. 11653003.  
BH is also partially supported by the Chinese National Youth Thousand Talents Program under the reference No. 110532102 and the Fundamental Research Funds for the Central Universities under the reference No.310421107. 
TJZ is supported by the National Science Foundation of China (Grants No. 11573006, 11929301), and National Key R$\&$D Program of China (2017YFA0402600).



\bibliographystyle{mnras}
\bibliography{MG_EC_refs} 

\begin{thebibliography}{}
\makeatletter
\relax
\def\mn@urlcharsother{\let\do\@makeother \do\$\do\&\do\#\do\^\do\_\do\%\do\~}
\def\mn@doi{\begingroup\mn@urlcharsother \@ifnextchar [ {\mn@doi@}
  {\mn@doi@[]}}
\def\mn@doi@[#1]#2{\def\@tempa{#1}\ifx\@tempa\@empty \href
  {http://dx.doi.org/#2} {doi:#2}\else \href {http://dx.doi.org/#2} {#1}\fi
  \endgroup}
\def\mn@eprint#1#2{\mn@eprint@#1:#2::\@nil}
\def\mn@eprint@arXiv#1{\href {http://arxiv.org/abs/#1} {{\tt arXiv:#1}}}
\def\mn@eprint@dblp#1{\href {http://dblp.uni-trier.de/rec/bibtex/#1.xml}
  {dblp:#1}}
\def\mn@eprint@#1:#2:#3:#4\@nil{\def\@tempa {#1}\def\@tempb {#2}\def\@tempc
  {#3}\ifx \@tempc \@empty \let \@tempc \@tempb \let \@tempb \@tempa \fi \ifx
  \@tempb \@empty \def\@tempb {arXiv}\fi \@ifundefined
  {mn@eprint@\@tempb}{\@tempb:\@tempc}{\expandafter \expandafter \csname
  mn@eprint@\@tempb\endcsname \expandafter{\@tempc}}}

\bibitem[\protect\citeauthoryear{Achitouv, Baldi, Puchwein  \& Weller}{Achitouv
  et~al.}{2016}]{Achitouv:2015yha}
Achitouv I.,  Baldi M.,  Puchwein E.,   Weller J.,  2016, \mn@doi [Phys. Rev.]
  {10.1103/PhysRevD.93.103522}, D93, 103522

\bibitem[\protect\citeauthoryear{{Barreira}, {Li}, {Hellwing}, {Lombriser},
  {Baugh}  \& {Pascoli}}{{Barreira} et~al.}{2014}]{2014JCAP...04..029B}
{Barreira} A.,  {Li} B.,  {Hellwing} W.~A.,  {Lombriser} L.,  {Baugh} C.~M.,
  {Pascoli} S.,  2014, \mn@doi [\jcap] {10.1088/1475-7516/2014/04/029}, \href
  {http://adsabs.harvard.edu/abs/2014JCAP...04..029B} {4, 029}

\bibitem[\protect\citeauthoryear{Bellini et~al.}{Bellini
  et~al.}{2018}]{Bellini:2017avd}
Bellini E.,  et~al., 2018, \mn@doi [Phys. Rev.] {10.1103/PhysRevD.97.023520},
  D97, 023520

\bibitem[\protect\citeauthoryear{Blas, Lesgourgues  \& Tram}{Blas
  et~al.}{2011}]{Blas:2011rf}
Blas D.,  Lesgourgues J.,   Tram T.,  2011, \mn@doi [JCAP]
  {10.1088/1475-7516/2011/07/034}, 1107, 034

\bibitem[\protect\citeauthoryear{Bloomfield, Flanagan, Park  \&
  Watson}{Bloomfield et~al.}{2013}]{Bloomfield:2012ff}
Bloomfield J.~K.,  Flanagan Ã.~Ã.,  Park M.,   Watson S.,  2013, \mn@doi [JCAP]
  {10.1088/1475-7516/2013/08/010}, 1308, 010

\bibitem[\protect\citeauthoryear{{Bond}, {Cole}, {Efstathiou}  \&
  {Kaiser}}{{Bond} et~al.}{1991}]{1991ApJ...379..440B}
{Bond} J.~R.,  {Cole} S.,  {Efstathiou} G.,   {Kaiser} N.,  1991, \mn@doi
  [\apj] {10.1086/170520}, \href
  {http://adsabs.harvard.edu/abs/1991ApJ...379..440B} {379, 440}

\bibitem[\protect\citeauthoryear{Bullock, Kolatt, Sigad, Somerville, Kravtsov,
  Klypin, Primack  \& Dekel}{Bullock et~al.}{2001}]{Bullock:1999he}
Bullock J.~S.,  Kolatt T.~S.,  Sigad Y.,  Somerville R.~S.,  Kravtsov A.~V.,
  Klypin A.~A.,  Primack J.~R.,   Dekel A.,  2001, \mn@doi [Mon. Not. Roy.
  Astron. Soc.] {10.1046/j.1365-8711.2001.04068.x}, 321, 559

\bibitem[\protect\citeauthoryear{{Burrage}, {Copeland}  \&
  {Stevenson}}{{Burrage} et~al.}{2015}]{2015PhRvD..91f5030B}
{Burrage} C.,  {Copeland} E.~J.,   {Stevenson} J.~A.,  2015, \mn@doi [\prd]
  {10.1103/PhysRevD.91.065030}, \href
  {http://adsabs.harvard.edu/abs/2015PhRvD..91f5030B} {91, 065030}

\bibitem[\protect\citeauthoryear{{Burrage}, {Copeland}, {Moss}  \&
  {Stevenson}}{{Burrage} et~al.}{2018}]{2018JCAP...01..056B}
{Burrage} C.,  {Copeland} E.~J.,  {Moss} A.,   {Stevenson} J.~A.,  2018,
  \mn@doi [\jcap] {10.1088/1475-7516/2018/01/056}, \href
  {http://adsabs.harvard.edu/abs/2018JCAP...01..056B} {1, 056}

\bibitem[\protect\citeauthoryear{{Carlberg} \& {Couchman}}{{Carlberg} \&
  {Couchman}}{1989}]{1989ApJ...340...47C}
{Carlberg} R.~G.,  {Couchman} H.~M.~P.,  1989, \mn@doi [\apj] {10.1086/167375},
  \href {http://adsabs.harvard.edu/abs/1989ApJ...340...47C} {340, 47}

\bibitem[\protect\citeauthoryear{Cataneo, Lombriser, Heymans, Mead, Barreira,
  Bose  \& Li}{Cataneo et~al.}{2019}]{Cataneo:2018cic}
Cataneo M.,  Lombriser L.,  Heymans C.,  Mead A.,  Barreira A.,  Bose S.,   Li
  B.,  2019, \mn@doi [Mon. Not. Roy. Astron. Soc.] {10.1093/mnras/stz1836},
  488, 2121

\bibitem[\protect\citeauthoryear{Cooray \& Hu}{Cooray \&
  Hu}{2001}]{Cooray:2000ry}
Cooray A.,  Hu W.,  2001, \mn@doi [Astrophys. J.] {10.1086/321376}, 554, 56

\bibitem[\protect\citeauthoryear{Cooray \& Sheth}{Cooray \&
  Sheth}{2002}]{Cooray:2002dia}
Cooray A.,  Sheth R.~K.,  2002, \mn@doi [Phys. Rept.]
  {10.1016/S0370-1573(02)00276-4}, 372, 1

\bibitem[\protect\citeauthoryear{{Dodelson}}{{Dodelson}}{2003}]{2003moco.book.....D}
{Dodelson} S.,  2003, {Modern cosmology}

\bibitem[\protect\citeauthoryear{{Doroshkevich}}{{Doroshkevich}}{1970}]{1970Ap......6..320D}
{Doroshkevich} A.~G.,  1970, \mn@doi [Astrophysics] {10.1007/BF01001625}, \href
  {http://adsabs.harvard.edu/abs/1970Ap......6..320D} {6, 320}

\bibitem[\protect\citeauthoryear{Dossett \& Ishak}{Dossett \&
  Ishak}{2012}]{Dossett:2012kd}
Dossett J.,  Ishak M.,  2012, \mn@doi [Phys. Rev.]
  {10.1103/PhysRevD.86.103008}, D86, 103008

\bibitem[\protect\citeauthoryear{Dossett, Ishak  \& Moldenhauer}{Dossett
  et~al.}{2011}]{Dossett:2011tn}
Dossett J.~N.,  Ishak M.,   Moldenhauer J.,  2011, \mn@doi [Phys. Rev.]
  {10.1103/PhysRevD.84.123001}, D84, 123001

\bibitem[\protect\citeauthoryear{{Efstathiou}, {Frenk}, {White}  \&
  {Davis}}{{Efstathiou} et~al.}{1988}]{1988MNRAS.235..715E}
{Efstathiou} G.,  {Frenk} C.~S.,  {White} S.~D.~M.,   {Davis} M.,  1988,
  \mn@doi [\mnras] {10.1093/mnras/235.3.715}, \href
  {http://adsabs.harvard.edu/abs/1988MNRAS.235..715E} {235, 715}

\bibitem[\protect\citeauthoryear{Espejo, Peirone, Raveri, Koyama, Pogosian  \&
  Silvestri}{Espejo et~al.}{2019}]{Espejo:2018hxa}
Espejo J.,  Peirone S.,  Raveri M.,  Koyama K.,  Pogosian L.,   Silvestri A.,
  2019, \mn@doi [Phys. Rev.] {10.1103/PhysRevD.99.023512}, D99, 023512

\bibitem[\protect\citeauthoryear{Frusciante \& Perenon}{Frusciante \&
  Perenon}{2019}]{Frusciante:2019xia}
Frusciante N.,  Perenon L.,  2019, Effective Field Theory of Dark Energy: a
  Review (\mn@eprint {arXiv} {1907.03150})

\bibitem[\protect\citeauthoryear{Frusciante, Papadomanolakis, Peirone  \&
  Silvestri}{Frusciante et~al.}{2019}]{Frusciante:2018vht}
Frusciante N.,  Papadomanolakis G.,  Peirone S.,   Silvestri A.,  2019, \mn@doi
  [JCAP] {10.1088/1475-7516/2019/02/029}, 1902, 029

\bibitem[\protect\citeauthoryear{Giocoli, Bartelmann, Sheth  \&
  Cacciato}{Giocoli et~al.}{2010}]{Giocoli:2010dm}
Giocoli C.,  Bartelmann M.,  Sheth R.~K.,   Cacciato M.,  2010, \mn@doi [Mon.
  Not. Roy. Astron. Soc.] {10.1111/j.1365-2966.2010.17108.x}, 408, 300

\bibitem[\protect\citeauthoryear{Gubitosi, Piazza  \& Vernizzi}{Gubitosi
  et~al.}{2013}]{Gubitosi:2012hu}
Gubitosi G.,  Piazza F.,   Vernizzi F.,  2013, \mn@doi [JCAP]
  {10.1088/1475-7516/2013/02/032}, 1302, 032

\bibitem[\protect\citeauthoryear{{Hinshaw} et~al.,}{{Hinshaw}
  et~al.}{2013}]{2013ApJS..208...19H}
{Hinshaw} G.,  et~al., 2013, \mn@doi [\apjs] {10.1088/0067-0049/208/2/19},
  \href {https://ui.adsabs.harvard.edu/abs/2013ApJS..208...19H} {208, 19}

\bibitem[\protect\citeauthoryear{Hojjati, Pogosian  \& Zhao}{Hojjati
  et~al.}{2011}]{Hojjati:2011ix}
Hojjati A.,  Pogosian L.,   Zhao G.-B.,  2011, \mn@doi [JCAP]
  {10.1088/1475-7516/2011/08/005}, 1108, 005

\bibitem[\protect\citeauthoryear{{Hu} \& {Sawicki}}{{Hu} \&
  {Sawicki}}{2007}]{2007PhRvD..76f4004H}
{Hu} W.,  {Sawicki} I.,  2007, \mn@doi [\prd] {10.1103/PhysRevD.76.064004},
  \href {http://adsabs.harvard.edu/abs/2007PhRvD..76f4004H} {76, 064004}

\bibitem[\protect\citeauthoryear{{Hu}, {Raveri}, {Frusciante}  \&
  {Silvestri}}{{Hu} et~al.}{2014}]{2014PhRvD..89j3530H}
{Hu} B.,  {Raveri} M.,  {Frusciante} N.,   {Silvestri} A.,  2014, \mn@doi
  [\prd] {10.1103/PhysRevD.89.103530}, \href
  {http://adsabs.harvard.edu/abs/2014PhRvD..89j3530H} {89, 103530}

\bibitem[\protect\citeauthoryear{Hu, Raveri, Rizzato  \& Silvestri}{Hu
  et~al.}{2016}]{Hu:2016zrh}
Hu B.,  Raveri M.,  Rizzato M.,   Silvestri A.,  2016, \mn@doi [Mon. Not. Roy.
  Astron. Soc.] {10.1093/mnras/stw775}, 459, 3880

\bibitem[\protect\citeauthoryear{{Hu}, {Liu}  \& {Cai}}{{Hu}
  et~al.}{2018}]{2018MNRAS.476L..65H}
{Hu} B.,  {Liu} X.-W.,   {Cai} R.-G.,  2018, \mn@doi [\mnras]
  {10.1093/mnrasl/sly032}, \href
  {http://adsabs.harvard.edu/abs/2018MNRAS.476L..65H} {476, L65}

\bibitem[\protect\citeauthoryear{Jenkins, Frenk, White, Colberg, Cole, Evrard,
  Couchman  \& Yoshida}{Jenkins et~al.}{2001}]{Jenkins:2000bv}
Jenkins A.,  Frenk C.~S.,  White S. D.~M.,  Colberg J.~M.,  Cole S.,  Evrard
  A.~E.,  Couchman H. M.~P.,   Yoshida N.,  2001, \mn@doi [Mon. Not. Roy.
  Astron. Soc.] {10.1046/j.1365-8711.2001.04029.x}, 321, 372

\bibitem[\protect\citeauthoryear{Khoury \& Weltman}{Khoury \&
  Weltman}{2004}]{PhysRevD.69.044026}
Khoury J.,  Weltman A.,  2004, \mn@doi [Phys. Rev. D]
  {10.1103/PhysRevD.69.044026}, 69, 044026

\bibitem[\protect\citeauthoryear{Kopp, Appleby, Achitouv  \& Weller}{Kopp
  et~al.}{2013}]{Kopp:2013lea}
Kopp M.,  Appleby S.~A.,  Achitouv I.,   Weller J.,  2013, \mn@doi [Phys. Rev.]
  {10.1103/PhysRevD.88.084015}, D88, 084015

\bibitem[\protect\citeauthoryear{{Lam} \& {Li}}{{Lam} \&
  {Li}}{2012}]{2012MNRAS.426.3260L}
{Lam} T.~Y.,  {Li} B.,  2012, \mn@doi [\mnras]
  {10.1111/j.1365-2966.2012.21746.x}, \href
  {http://adsabs.harvard.edu/abs/2012MNRAS.426.3260L} {426, 3260}

\bibitem[\protect\citeauthoryear{{Lam} \& {Sheth}}{{Lam} \&
  {Sheth}}{2008}]{2008MNRAS.386..407L}
{Lam} T.~Y.,  {Sheth} R.~K.,  2008, \mn@doi [\mnras]
  {10.1111/j.1365-2966.2008.13038.x}, \href
  {http://adsabs.harvard.edu/abs/2008MNRAS.386..407L} {386, 407}

\bibitem[\protect\citeauthoryear{Lewis, Challinor  \& Lasenby}{Lewis
  et~al.}{2000}]{Lewis:1999bs}
Lewis A.,  Challinor A.,   Lasenby A.,  2000, \mn@doi [Astrophys. J.]
  {10.1086/309179}, 538, 473

\bibitem[\protect\citeauthoryear{{Li} \& {Efstathiou}}{{Li} \&
  {Efstathiou}}{2012}]{2012MNRAS.421.1431L}
{Li} B.,  {Efstathiou} G.,  2012, \mn@doi [\mnras]
  {10.1111/j.1365-2966.2011.20404.x}, \href
  {http://adsabs.harvard.edu/abs/2012MNRAS.421.1431L} {421, 1431}

\bibitem[\protect\citeauthoryear{{Li} \& {Lam}}{{Li} \&
  {Lam}}{2012}]{2012MNRAS.425..730L}
{Li} B.,  {Lam} T.~Y.,  2012, \mn@doi [\mnras]
  {10.1111/j.1365-2966.2012.21592.x}, \href
  {http://adsabs.harvard.edu/abs/2012MNRAS.425..730L} {425, 730}

\bibitem[\protect\citeauthoryear{{Li}, {Zhao}, {Teyssier}  \& {Koyama}}{{Li}
  et~al.}{2012}]{2012JCAP...01..051L}
{Li} B.,  {Zhao} G.-B.,  {Teyssier} R.,   {Koyama} K.,  2012, \mn@doi [\jcap]
  {10.1088/1475-7516/2012/01/051}, \href
  {https://ui.adsabs.harvard.edu/abs/2012JCAP...01..051L} {2012, 051}

\bibitem[\protect\citeauthoryear{{Li}, {Hellwing}, {Koyama}, {Zhao}, {Jennings}
   \& {Baugh}}{{Li} et~al.}{2013a}]{2013MNRAS.428..743L}
{Li} B.,  {Hellwing} W.~A.,  {Koyama} K.,  {Zhao} G.-B.,  {Jennings} E.,
  {Baugh} C.~M.,  2013a, \mn@doi [\mnras] {10.1093/mnras/sts072}, \href
  {http://adsabs.harvard.edu/abs/2013MNRAS.428..743L} {428, 743}

\bibitem[\protect\citeauthoryear{{Li}, {Zhao}  \& {Koyama}}{{Li}
  et~al.}{2013b}]{2013JCAP...05..023L}
{Li} B.,  {Zhao} G.-B.,   {Koyama} K.,  2013b, \mn@doi [\jcap]
  {10.1088/1475-7516/2013/05/023}, \href
  {https://ui.adsabs.harvard.edu/abs/2013JCAP...05..023L} {2013, 023}

\bibitem[\protect\citeauthoryear{{Lombriser}}{{Lombriser}}{2014}]{2014AnP...526..259L}
{Lombriser} L.,  2014, \mn@doi [Annalen der Physik] {10.1002/andp.201400058},
  \href {http://adsabs.harvard.edu/abs/2014AnP...526..259L} {526, 259}

\bibitem[\protect\citeauthoryear{Lombriser}{Lombriser}{2016}]{Lombriser:2016zfz}
Lombriser L.,  2016, \mn@doi [JCAP] {10.1088/1475-7516/2016/11/039}, 1611, 039

\bibitem[\protect\citeauthoryear{{Lombriser}, {Li}, {Koyama}  \&
  {Zhao}}{{Lombriser} et~al.}{2013a}]{2013PhRvD..87l3511L}
{Lombriser} L.,  {Li} B.,  {Koyama} K.,   {Zhao} G.-B.,  2013a, \mn@doi [\prd]
  {10.1103/PhysRevD.87.123511}, \href
  {http://adsabs.harvard.edu/abs/2013PhRvD..87l3511L} {87, 123511}

\bibitem[\protect\citeauthoryear{Lombriser, Li, Koyama  \& Zhao}{Lombriser
  et~al.}{2013b}]{Lombriser:2013wta}
Lombriser L.,  Li B.,  Koyama K.,   Zhao G.-B.,  2013b, \mn@doi [Phys. Rev.]
  {10.1103/PhysRevD.87.123511}, D87, 123511

\bibitem[\protect\citeauthoryear{{Lombriser}, {Koyama}  \& {Li}}{{Lombriser}
  et~al.}{2014}]{2014JCAP...03..021L}
{Lombriser} L.,  {Koyama} K.,   {Li} B.,  2014, \mn@doi [\jcap]
  {10.1088/1475-7516/2014/03/021}, \href
  {http://adsabs.harvard.edu/abs/2014JCAP...03..021L} {3, 021}

\bibitem[\protect\citeauthoryear{{Maggiore} \& {Riotto}}{{Maggiore} \&
  {Riotto}}{2010}]{2010ApJ...717..515M}
{Maggiore} M.,  {Riotto} A.,  2010, \mn@doi [\apj]
  {10.1088/0004-637X/717/1/515}, \href
  {http://adsabs.harvard.edu/abs/2010ApJ...717..515M} {717, 515}

\bibitem[\protect\citeauthoryear{Mead, Peacock, Heymans, Joudaki  \&
  Heavens}{Mead et~al.}{2015}]{Mead:2015yca}
Mead A.,  Peacock J.,  Heymans C.,  Joudaki S.,   Heavens A.,  2015, \mn@doi
  [Mon. Not. Roy. Astron. Soc.] {10.1093/mnras/stv2036}, 454, 1958

\bibitem[\protect\citeauthoryear{{Mo}, {van den Bosch}  \& {White}}{{Mo}
  et~al.}{2010}]{2010gfe..book.....M}
{Mo} H.,  {van den Bosch} F.~C.,   {White} S.,  2010, {Galaxy Formation and
  Evolution}

\bibitem[\protect\citeauthoryear{Monaco, Theuns  \& Taffoni}{Monaco
  et~al.}{2002a}]{Monaco:2001jg}
Monaco P.,  Theuns T.,   Taffoni G.,  2002a, \mn@doi [Mon. Not. Roy. Astron.
  Soc.] {10.1046/j.1365-8711.2002.05162.x}, 331, 587

\bibitem[\protect\citeauthoryear{Monaco, Theuns, Taffoni, Governato, Quinn  \&
  Stadel}{Monaco et~al.}{2002b}]{Monaco:2001jf}
Monaco P.,  Theuns T.,  Taffoni G.,  Governato F.,  Quinn T.~R.,   Stadel J.,
  2002b, \mn@doi [Astrophys. J.] {10.1086/324182}, 564, 8

\bibitem[\protect\citeauthoryear{Monaco, Sefusatti, Borgani, Crocce, Fosalba,
  Sheth  \& Theuns}{Monaco et~al.}{2013}]{Monaco:2013qta}
Monaco P.,  Sefusatti E.,  Borgani S.,  Crocce M.,  Fosalba P.,  Sheth R.~K.,
  Theuns T.,  2013, \mn@doi [Mon. Not. Roy. Astron. Soc.]
  {10.1093/mnras/stt907}, 433, 2389

\bibitem[\protect\citeauthoryear{More, Kravtsov, Dalal  \& Gottlober}{More
  et~al.}{2011}]{More:2011dc}
More S.,  Kravtsov A.,  Dalal N.,   Gottlober S.,  2011, \mn@doi [Astrophys. J.
  Suppl.] {10.1088/0067-0049/195/1/4}, 195, 4

\bibitem[\protect\citeauthoryear{{Oyaizu}, {Lima}  \& {Hu}}{{Oyaizu}
  et~al.}{2008}]{2008PhRvD..78l3524O}
{Oyaizu} H.,  {Lima} M.,   {Hu} W.,  2008, \mn@doi [\prd]
  {10.1103/PhysRevD.78.123524}, \href
  {http://adsabs.harvard.edu/abs/2008PhRvD..78l3524O} {78, 123524}

\bibitem[\protect\citeauthoryear{Pace, Manera, Bacon, Crittenden  \&
  Percival}{Pace et~al.}{2015}]{Pace:2015nda}
Pace F.,  Manera M.,  Bacon D.~J.,  Crittenden R.,   Percival W.~J.,  2015,
  \mn@doi [Mon. Not. Roy. Astron. Soc.] {10.1093/mnras/stv2019}, 454, 708

\bibitem[\protect\citeauthoryear{Peirone, Koyama, Pogosian, Raveri  \&
  Silvestri}{Peirone et~al.}{2018}]{Peirone:2017ywi}
Peirone S.,  Koyama K.,  Pogosian L.,  Raveri M.,   Silvestri A.,  2018,
  \mn@doi [Phys. Rev.] {10.1103/PhysRevD.97.043519}, D97, 043519

\bibitem[\protect\citeauthoryear{Piazza \& Vernizzi}{Piazza \&
  Vernizzi}{2013}]{Piazza:2013coa}
Piazza F.,  Vernizzi F.,  2013, \mn@doi [Class. Quant. Grav.]
  {10.1088/0264-9381/30/21/214007}, 30, 214007

\bibitem[\protect\citeauthoryear{{Press} \& {Schechter}}{{Press} \&
  {Schechter}}{1974}]{1974ApJ...187..425P}
{Press} W.~H.,  {Schechter} P.,  1974, \mn@doi [\apj] {10.1086/152650}, \href
  {http://adsabs.harvard.edu/abs/1974ApJ...187..425P} {187, 425}

\bibitem[\protect\citeauthoryear{{Prunet}, {Pichon}, {Aubert}, {Pogosyan},
  {Teyssier}  \& {Gottloeber}}{{Prunet} et~al.}{2008}]{2008ApJS..178..179P}
{Prunet} S.,  {Pichon} C.,  {Aubert} D.,  {Pogosyan} D.,  {Teyssier} R.,
  {Gottloeber} S.,  2008, \mn@doi [\apjs] {10.1086/590370}, \href
  {https://ui.adsabs.harvard.edu/abs/2008ApJS..178..179P} {178, 179}

\bibitem[\protect\citeauthoryear{{Raveri}, {Hu}, {Frusciante}  \&
  {Silvestri}}{{Raveri} et~al.}{2014}]{2014PhRvD..90d3513R}
{Raveri} M.,  {Hu} B.,  {Frusciante} N.,   {Silvestri} A.,  2014, \mn@doi
  [\prd] {10.1103/PhysRevD.90.043513}, \href
  {http://adsabs.harvard.edu/abs/2014PhRvD..90d3513R} {90, 043513}

\bibitem[\protect\citeauthoryear{{Schmidt}, {Lima}, {Oyaizu}  \&
  {Hu}}{{Schmidt} et~al.}{2009a}]{2009PhRvD..79h3518S}
{Schmidt} F.,  {Lima} M.,  {Oyaizu} H.,   {Hu} W.,  2009a, \mn@doi [\prd]
  {10.1103/PhysRevD.79.083518}, \href
  {http://adsabs.harvard.edu/abs/2009PhRvD..79h3518S} {79, 083518}

\bibitem[\protect\citeauthoryear{Schmidt, Lima, Oyaizu  \& Hu}{Schmidt
  et~al.}{2009b}]{Schmidt:2008tn}
Schmidt F.,  Lima M.~V.,  Oyaizu H.,   Hu W.,  2009b, \mn@doi [Phys. Rev.]
  {10.1103/PhysRevD.79.083518}, D79, 083518

\bibitem[\protect\citeauthoryear{Sheth \& Tormen}{Sheth \&
  Tormen}{1999}]{Sheth:1999mn}
Sheth R.~K.,  Tormen G.,  1999, \mn@doi [Mon. Not. Roy. Astron. Soc.]
  {10.1046/j.1365-8711.1999.02692.x}, 308, 119

\bibitem[\protect\citeauthoryear{{Sheth} \& {Tormen}}{{Sheth} \&
  {Tormen}}{2002}]{2002MNRAS.329...61S}
{Sheth} R.~K.,  {Tormen} G.,  2002, \mn@doi [\mnras]
  {10.1046/j.1365-8711.2002.04950.x}, \href
  {http://adsabs.harvard.edu/abs/2002MNRAS.329...61S} {329, 61}

\bibitem[\protect\citeauthoryear{{Sheth}, {Mo}  \& {Tormen}}{{Sheth}
  et~al.}{2001}]{2001MNRAS.323....1S}
{Sheth} R.~K.,  {Mo} H.~J.,   {Tormen} G.,  2001, \mn@doi [\mnras]
  {10.1046/j.1365-8711.2001.04006.x}, \href
  {http://adsabs.harvard.edu/abs/2001MNRAS.323....1S} {323, 1}

\bibitem[\protect\citeauthoryear{Taffoni, Monaco  \& Theuns}{Taffoni
  et~al.}{2002}]{Taffoni:2001jh}
Taffoni G.,  Monaco P.,   Theuns T.,  2002, \mn@doi [Mon. Not. Roy. Astron.
  Soc.] {10.1046/j.1365-8711.2002.05441.x}, 333, 623

\bibitem[\protect\citeauthoryear{Tassev, Zaldarriaga  \& Eisenstein}{Tassev
  et~al.}{2013}]{Tassev:2013pn}
Tassev S.,  Zaldarriaga M.,   Eisenstein D.,  2013, \mn@doi [JCAP]
  {10.1088/1475-7516/2013/06/036}, 1306, 036

\bibitem[\protect\citeauthoryear{{Teyssier}}{{Teyssier}}{2002}]{2002A&A...385..337T}
{Teyssier} R.,  2002, \mn@doi [\aap] {10.1051/0004-6361:20011817}, \href
  {https://ui.adsabs.harvard.edu/abs/2002A&A...385..337T} {385, 337}

\bibitem[\protect\citeauthoryear{Valogiannis \& Bean}{Valogiannis \&
  Bean}{2017}]{Valogiannis:2016ane}
Valogiannis G.,  Bean R.,  2017, \mn@doi [Phys. Rev.]
  {10.1103/PhysRevD.95.103515}, D95, 103515

\bibitem[\protect\citeauthoryear{Valogiannis \& Bean}{Valogiannis \&
  Bean}{2019}]{Valogiannis:2019xed}
Valogiannis G.,  Bean R.,  2019, \mn@doi [Phys. Rev.]
  {10.1103/PhysRevD.99.063526}, D99, 063526

\bibitem[\protect\citeauthoryear{Winther et~al.}{Winther
  et~al.}{2015}]{Winther:2015wla}
Winther H.~A.,  et~al., 2015, \mn@doi [Mon. Not. Roy. Astron. Soc.]
  {10.1093/mnras/stv2253}, 454, 4208

\bibitem[\protect\citeauthoryear{Winther, Koyama, Manera, Wright  \&
  Zhao}{Winther et~al.}{2017}]{Winther:2017jof}
Winther H.~A.,  Koyama K.,  Manera M.,  Wright B.~S.,   Zhao G.-B.,  2017,
  \mn@doi [JCAP] {10.1088/1475-7516/2017/08/006}, 1708, 006

\bibitem[\protect\citeauthoryear{Winther, Casas, Baldi, Koyama, Li, Lombriser
  \& Zhao}{Winther et~al.}{2019}]{Winther:2019mus}
Winther H.,  Casas S.,  Baldi M.,  Koyama K.,  Li B.,  Lombriser L.,   Zhao
  G.-B.,  2019, Emulators for the non-linear matter power spectrum beyond
  $Λ$CDM (\mn@eprint {arXiv} {1903.08798})

\bibitem[\protect\citeauthoryear{{Zel'dovich}}{{Zel'dovich}}{1970}]{1970A&A.....5...84Z}
{Zel'dovich} Y.~B.,  1970, \aap, \href
  {http://adsabs.harvard.edu/abs/1970A%26A.....5...84Z} {5, 84}

\bibitem[\protect\citeauthoryear{{Zhang} \& {Hui}}{{Zhang} \&
  {Hui}}{2006}]{2006ApJ...641..641Z}
{Zhang} J.,  {Hui} L.,  2006, \mn@doi [\apj] {10.1086/499802}, \href
  {http://adsabs.harvard.edu/abs/2006ApJ...641..641Z} {641, 641}

\bibitem[\protect\citeauthoryear{Zhao}{Zhao}{2014}]{Zhao:2013dza}
Zhao G.-B.,  2014, \mn@doi [Astrophys. J. Suppl.] {10.1088/0067-0049/211/2/23},
  211, 23

\bibitem[\protect\citeauthoryear{Zhao, Pogosian, Silvestri  \& Zylberberg}{Zhao
  et~al.}{2009}]{Zhao:2008bn}
Zhao G.-B.,  Pogosian L.,  Silvestri A.,   Zylberberg J.,  2009, \mn@doi [Phys.
  Rev.] {10.1103/PhysRevD.79.083513}, D79, 083513

\bibitem[\protect\citeauthoryear{Zucca, Pogosian, Silvestri  \& Zhao}{Zucca
  et~al.}{2019}]{Zucca:2019xhg}
Zucca A.,  Pogosian L.,  Silvestri A.,   Zhao G.-B.,  2019, \mn@doi [JCAP]
  {10.1088/1475-7516/2019/05/001}, 1905, 001

\bibitem[\protect\citeauthoryear{Zumalacárregui, Bellini, Sawicki, Lesgourgues
   \& Ferreira}{Zumalacárregui et~al.}{2017}]{Zumalacarregui:2016pph}
Zumalacárregui M.,  Bellini E.,  Sawicki I.,  Lesgourgues J.,   Ferreira
  P.~G.,  2017, \mn@doi [JCAP] {10.1088/1475-7516/2017/08/019}, 1708, 019

\bibitem[\protect\citeauthoryear{van Daalen \& Schaye}{van Daalen \&
  Schaye}{2015}]{vanDaalen:2015msa}
van Daalen M.~P.,  Schaye J.,  2015, \mn@doi [Mon. Not. Roy. Astron. Soc.]
  {10.1093/mnras/stv1456}, 452, 2247

\makeatother
\end{thebibliography}







\bsp	
\label{lastpage}
\end{document}